\documentclass[12pt]{iopart}

\usepackage{xcolor}
\usepackage{graphicx}
\usepackage{cite}
\usepackage[linktoc=all,backref=page]{hyperref}
\hypersetup{colorlinks,allcolors=blue}
\hypersetup{linktocpage}
\usepackage{iopams}
\usepackage{mathtools}
\usepackage{amssymb}
\usepackage{longtable}
\usepackage{physics}
\usepackage{lmodern}

\newcommand{\expect}[1]{\langle #1 \rangle}

\makeatletter
\newcommand*\mysize{%
  \@setfontsize\mysize{6.5}{8.5}%
}
\makeatother

\begin{document}

\title[]{Many-body Physics of Ultracold Alkaline-Earth atoms with SU($N$)-symmetric interactions}

\author{Eduardo Ibarra-Garc\'ia-Padilla$^{1,2}$, Sayan Choudhury$^3$}
\address{$^1$ Department of Physics and Astronomy, University of California,  Davis, CA 95616, USA}
\address{$^2$ Department of Physics and Astronomy, San Jos\'e State University, San Jos\'e, California 95192, USA}
\address{$^3$ Harish-Chandra Research Institute, a CI of Homi Bhabha National Institute, Chhatnag Road, Jhunsi, Allahabad 211019}
\ead{edibarra@ucdavis.edu, sayanchoudhury@hri.res.in}
\vspace{10pt}
% \begin{indented}
% \item[] \today
% \end{indented}

\begin{abstract}
Symmetries play a crucial role in understanding phases of matter and the transitions between them. Theoretical investigations of quantum models with SU($N$) symmetry have provided important insights into many-body phenomena. However, these models have generally remained a theoretical idealization, since it is very difficult to exactly realize the SU($N$) symmetry in conventional quantum materials for large $N$. Intriguingly however, in recent years, ultracold alkaline-earth-atom (AEA) quantum simulators have paved the path to realize SU($N$)-symmetric many-body models, where $N$ is tunable and can be as large as 10. This symmetry emerges due to the closed shell structure of AEAs, thereby leading to a perfect decoupling of the electronic degrees of freedom from the nuclear spin. In this work, we provide a systematic review of recent theoretical and experimental work on the many-body physics of these systems. We first discuss the thermodynamic properties and collective modes of trapped Fermi gases, highlighting the enhanced interaction effects that appear as $N$ increases. We then discuss the properties of the SU($N$) Fermi-Hubbard model, focusing on some of the major experimental achievements in this area. We conclude with a compendium highlighting some of the significant theoretical progress on SU($N$) lattice models and a discussion of some exciting directions for future research.
\end{abstract}

\vspace{1pc}
\vspace{1pc}

%\submitto{\JPCM}

\maketitle
\vspace{1em}

\section{Introduction}
Rapid advances in the development of ultracold atomic systems have provided physicists with a powerful platform to explore various facets of many-body physics~\cite{lewenstein2012ultracold}. In particular, ultracold atoms loaded in optical lattices provide a versatile platform for the quantum simulation of both equilibrium and non-equilibrium physics~\cite{windpassinger2013engineering,choi2023quantum}. These systems provide unprecedented control over the effective lattice geometries as well as the strength and range of the inter-atomic interactions, thereby providing a pathway to address unresolved questions about emergent phenomena~\cite{goldman2016topological,windpassinger2015specific,arguello2024synthetic,impertro2023local}. Furthermore, these systems can be employed to synthesize new forms of quantum matter that go beyond the capabilities of conventional quantum materials and can be harnessed for quantum information processing tasks~\cite{wang2016single,kumar2018sorting,young2022tweezer}.\\

Most initial experimental efforts in this area focused on trapping and cooling alkali atoms~\cite{bloch2005ultracold,bloch2008many}. These atoms have only one valence electron thereby enabling the application of several quantum control techniques. Furthermore, in certain atomic species such as $^{6}$Li or $^{40}$K, the scattering length can be tuned by employing magnetic Feshbach resonances, thereby providing a knob to control the strength of the interactions~\cite{kokkelmans2015feshbach}. However, the range of phenomena that can be investigated using alkali atoms is somewhat limited due to their atomic structure. A promising avenue to go beyond the limitations of alkali gases has been provided by Alkaline-Earth-like atoms (AEAs)~\cite{He_2019}. AEAs are the elements that belong to group II of the periodic table (Be, Mg, C, Sr, Ba, and Ra), but also the rare-earth element Yb. These elements are characterized by full inner shells and two outer valence electrons in a filled s-shell. Due to their filled inner shells, the two valence electrons govern most of these atoms' chemical and electronic properties. Most experiments with AEAs in have employed $^{173}$Yb ($I=5/2$) and $^{87}$Sr ($I=9/2$); for concreteness, we will discuss the electronic structure of Yb in detail in the next sub-section, but the ideas are applicable also for Sr.\\

\subsection{Electronic structure of AEAs}

The electronic configuration of $^{173}$Yb is [Xe]4f$^{14}$6s$^2$. It has filled f- and s-shells. Yb exhibits an electronic structure with spin-singlet ($S=0$) and spin-triplet ($S=1$) manifolds. To a first approximation, its level structure is well described in the LS-coupling scheme, where coupling between the total orbital angular momentum \textbf{L} and the total spin of the valence electrons \textbf{S} gives the total electronic angular momentum \textbf{J = L + S}, and eigenstates are labelled as $^{2S+1}L_J$. Fig.~\ref{fig::Yb_electronic} displays the level structure for $^{173}$Yb with the most significant optical transitions. The ground state of $^{173}$Yb is a spin singlet $^1S_0$ with $J=0$; it also exhibits two metastable states, the $^3P_0$ state and the $^3P_2$ state. In particular, the $^3P_0$ state has a lifetime $\tau \approx 20$ seconds, and also has no total electronic angular momentum ($J=0$) like the ground state $^1S_0$. This has far-reaching implications on the properties of the states, since in this case:

\begin{figure}[htbp!]
\centering
\includegraphics[width=0.9\linewidth]{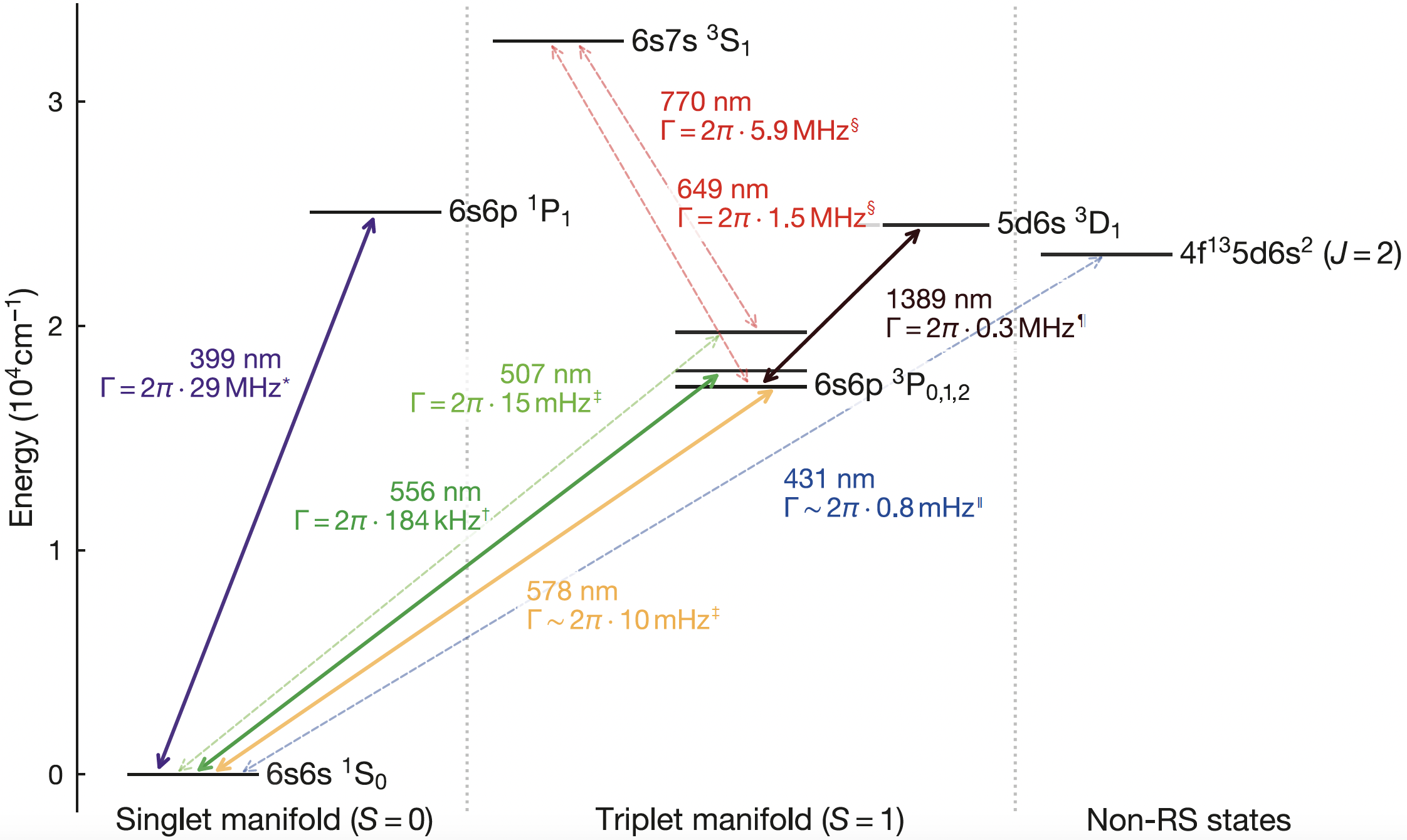}
\caption{Partial electronic structure of ytterbium. The states are labelled according to the Russell-Saunders (RS) notation when the f-shell is closed and in terms of total angular momentum $J$ when the f-shell is open. The wavelength $\lambda$ and the linewidth $\Gamma$ are specified for each transition. References for the values of the linewidths: $^*$~\cite{Takasu2004}, $^\dag$~\cite{Pandey2009}, $^\ddag$~\cite{Porsev2004}, $^\S$~\cite{Cho2012}, $^\P$~\cite{Beloy2012}, \textsuperscript{\textbardbl}~\cite{Safronova2018,Dzuba2018}. Figure and caption reprinted with copyright permission of Ref.~\cite{pasqualettithesis}.}
\label{fig::Yb_electronic} 
\end{figure}

\begin{itemize}
    \item The total angular momentum of the atom $\textbf{F}$ is given only by the nuclear spin $\textbf{I}$ ($\textbf{F=I}$). The SU($N$) symmetric nature of interactions for the fermionic isotopes results from this decoupling of the nuclear spin degree of freedom from the electronic structure, as we will discuss in the following section~\footnote{Yb has fermionic isotopes $^{171}$Yb and $^{173}$Yb, which possess a nuclear spin $I=1/2$ and $I=5/2$, respectively; and bosonic isotopes $^{168}$Yb, $^{170}$Yb, $^{172}$Yb, $^{174}$Yb, and $^{176}$Yb, all which have $I=0$.}. 
    
    \item The state does not exhibit a hyperfine structure, since $\mathbf{I} \cdot \mathbf{J} = 0$, and therefore is almost completely insensitive to magnetic fields because fermionic and bosonic isotopes have a weak or zero nuclear magnetic moment, respectively~\footnote{For fermions, the only magnetic moment arises from the nuclear spin, which is essentially irrelevant since the nuclear magneton $\mu_N$ is approximately 2000 times smaller than the Bohr magneton $\mu_B$ of the electron.}. This insensitivity has relevant consequences in how interactions can be tuned for $^{173}$Yb atoms, since magnetic Feshbach resonances, which are typically employed to control the s-wave scattering length between two hyperfine states in alkali-atoms~\cite{Pethick_book}, are experimentally unavailable due to the requirement for extremely high magnetic fields that are not achievable in laboratories. There are, however, optical Feshbach resonances that can be used to tune interactions between different $m_I$ states. These have been used by several groups to realize Optical Stern Gerlach measurements and/or break the SU($N$) symmetric interactions~\cite{scazzathesis,scazza2014,Huang2020}.
\end{itemize}

\subsection{SU(N)-symmetric interactions}

Since $J=0$, the ground state $^1S_0$ and the metastable state $^3P_0$ exhibit perfect decoupling of the nuclear spin from the electronic structure due to the lack of hyperfine structure. In these states, the spin is protected inside the nucleus and it is not affected by the physics occurring at the electronic cloud distance scales. This has an important effect on the properties of atomic collisions because, aside from Pauli exclusion, nuclei can only affect collisions through hyperfine coupling to the electron angular momentum~\footnote{Magnetic dipole-dipole interactions between the nuclear spin of two atoms is negligible in comparison to the induced dipole-dipole interactions of the electronic clouds~\cite{Gorshkov2010}.}. In dilute atomic gases at low-temperatures, scattering properties are well characterized by the $s$-wave scattering length. For fermionic gases with $N=2I+1$ spin components is possible to model the collisions with the following pseudo-potential~\cite{Yip1999},
\begin{equation}\label{eq:pseudo_potential}
    V(\textbf{r}) =  \frac{4 \pi \hbar^2}{m} \sum_{F_t =0,2, \dots}^{N -2} a_{F_t}\delta(\textbf{r}) \mathcal{P}_{F_t},
\end{equation}
where $\hbar$ is the reduced Planck's constant, $m$ is the mass of the colliding atoms, $\mathcal{P}_{F_t}$ is the projector on states with even total spin $F_t = 0,2,4,\dots,N-2$ of the atom pair, and $a_{F_t}$ is the scattering length for a given $F_t$~\footnote{Only states with even total spin $F_t$ can contribute to the scattering process because of the antisymmetrization of the wavefunction. In $s$-wave scattering collisions the spatial wave function is symmetric, and therefore the spin wave function has to be antisymmetric.}.\\

For the two-body collision of atoms in states $\vert F, m_1 \rangle$, $\vert F, m_2 \rangle$ and total spin $\vert F_t, m_t =  m_1 + m_2\rangle$, the initial state $| F,m_1\rangle | F,m_2\rangle$ will couple to a different spin combination $| F,m_3\rangle | F,m_4\rangle$ via the pseudo-potential in eq.~\eqref{eq:pseudo_potential},
\begin{equation}\label{eq:spin_changing_collisions}
    \langle F,m_4 ; F,m_3| V(\textbf{r}) | F,m_1 ; F,m_2\rangle= \frac{4 \pi \hbar^2}{m} \delta(\mathbf{r}) \sum_{F_t =0,2, \dots}^{N -2} \sum_{m_t = -F_t}^{F_t} a_{F_t} C_{ F m_1 F m_2}^{F_t m_t} C_{F m_3 F m_4}^{F_t m_t}.
\end{equation}
where $C_{j_1 m_1 j_2 m_2}^{JM}$ are the Clebsh-Gordan coefficients and we used that $\mathcal{P}_{F_t} =  \sum_{m_t = - F_t}^{F_t} | F_t, m_t \rangle \langle F_t, m_t|$. Eq.~\eqref{eq:spin_changing_collisions} reflects that momentum conservation ensures that the total spin $F_t$ and its projection $m_t$ are conserved during the collision ($m_t = m_1 + m_2 = m_3 + m_4$), but the spin projection of the individual atoms is not. \\

In the case of AEAs, the scattering lengths are equal for all possible $F_t$ pairs ($a_{F_t} = a \, \forall F_t$), since the nuclear spin is decoupled from the electronic structure, and therefore its influence in the scattering process is simply reduced to Pauli exclusion principle. Because of the orthogonality relationships of the Clebsch-Gordan coefficients $\sum_J \sum_M C^{JM}_{j_1m_2j_2m_2}C^{JM}_{j_1m_{1'}j_2m_{2'}} = \delta_{m_1,m_{1'}}\delta_{m_2,m_{2'}}$ one observes that in contrast to the general case of collisions, for AEAs the spin projection $m_F$ of each colliding atom is preserved and thus spin relaxation to other $m_F$ states is forbidden. This means that the interaction will be SU($N$) symmetric, and the interaction pseudopotential simplifies to $ V(\textbf{r}) =  (4 \pi \hbar^2/m) a\delta(\textbf{r})$ for all possible pairs of spin projections. Ref.~\cite{Gorshkov2010} provides theoretical estimates for the SU($N$) symmetry breaking of AEAs. The variation in the scattering length for different nuclear spins in the ground state $^1S_0$ is of the order $\delta a_{gg}/a_{gg} \sim 10^{-9}$, while for the excited metastable state $^3P_0$ these are of order $\delta a_{ee}/a_{ee} \sim \delta a_{eg}^\pm/a_{eg}^\pm \sim 10^{-3}$~\footnote{Scattering processes between two atoms in the ground state are denoted by $a_{gg}$, two atoms in the excited state by $a_{ee}$, and one atom in the excited state and the other one in the ground state in their triplet (+) or singlet (-) configuration by $a_{eg}^{\pm}$.} (here the perfect decoupling is slightly broken by the admixture with higher-lying $P$ states with $J \neq 0$). 

\subsection{Structure of the review}
Now that we have introduced the electronic structure of AEA gases and demonstrated that they exhibit an emergent SU($N$) symmetry, we will proceed to review the many-body physics of these systems. This review is organized as follows.  In sec.~\ref{sec:trapped}, we discuss some interesting theoretical advancements and experimental results on trapped SU($N$) Fermi gases. In sec.~\ref{sec:OL}, we discuss some of the major works on quantum simulation of the SU($N$) Fermi-Hubbard model with ultracold AEAs in optical lattices. While we primarily focus on experimental achievements, we also provide a compendium of theoretical works on SU($N$) lattice models such as the Hubbard model, the Heisenberg model, and the $t-J$ model. Furthermore, we note that while lattice AEA systems provide a powerful platform for precision timekeeping and quantum computing, we do not delve into these aspects in this review. We conclude in sec.~\ref{sec:Future} by outlining some interesting directions for future research. \\

\section{Many-body physics of Trapped AEA Gases}
\label{sec:trapped}
The enlarged SU($N$) symmetry results in enhanced interaction effects in trapped AEA gases. This has remarkable consequences such as a strong $N$-dependence of the compressibility and collective mode frequencies as well as bosonization in higher dimensions. We now proceed to expand on these developments.

\subsection{Thermodynamics of the SU(N) Fermi Liquid}
We begin by examining the extension of the Fermi liquid theory for $N-$component fermions with SU($N$) symmetric interactions; this system is described by:
\begin{eqnarray}
    H &=& \sum_{{\bf k}, \gamma} \left(\frac{k^2}{2m} - \mu \right) c_{{\bf k}, \gamma}^{\dagger}c_{{\bf k}, \gamma} \nonumber \\
    &+& \frac{g}{2} \sum_{{\bf k_1},{\bf k_2},{\bf k_3},{\bf k_4} \gamma_1 \ne \gamma_2} c_{{\bf k_1}, \gamma_1}^{\dagger} c_{{\bf k_2}, \gamma_2}^{\dagger} c_{{\bf k_3}, \gamma_1} c_{{\bf k_4}, \gamma_2} \delta_{{\bf k_1 + k_2},{\bf k_3 + k_4}},
\end{eqnarray}
where $c_{{\bf k}, \gamma}$ is the annihilation operator for a fermion of species $\gamma$ and momentum ${\bf k}$. Before discussing the properties of the SU($N$) Fermi liquid, let us recall the standard Fermi Liquid (FL) theory formulated for a two-component fermionic system with SU(2) symmetry. We note that when interactions are adiabatically switched on in a Fermi gas, the one-particle states continuously evolve into quasi-particle states with the same spin and charge. For a spinless system, these quasi-particles are described by a distribution function $n({\bf k}) = n_0({\bf k}) + \delta n({\bf k})$, where $n_0({\bf k})$ is the distribution function of the non-interacting fermions. The change in energy, $\delta E$ due to a change in the distribution function $\delta n$ is given by:
\begin{equation}
    \delta E = V \int \frac{d^3 k}{(2 \pi)^3} \epsilon({\bf k}) \delta n({\bf k}),
\end{equation}
where $\epsilon({\bf k})$ is the quasi-particle energy. Furthermore, the change in the quasi-particle energy $\delta \epsilon ({\bf k})$ due to a change in the distribution function, $\delta n ({\bf k})$ is given by:
\begin{equation}
   \delta \epsilon({\bf k}) =  \sum_{{\bf k^{\prime}}} f({\bf k, k^{\prime}})\delta n({\bf k^{\prime}}),
\end{equation}
where $ f({\bf k, k^{\prime}}) = f({\bf k^{\prime},k})$.\\

Following Lifshitz and Pitaveskii, we can extend this treatment to a spin-$1/2$ Fermi gas~\cite{lifshitz1980statistical}. In this case, the distribution function, $n({\bf k})$ becomes a $2 \times 2$ matrix in terms of the spin variables, such that:
\begin{equation}
  N_0 =  V \sum_{\alpha} \int \frac{d^3 k}{ (2 \pi)^3} n_{\alpha, \alpha} ({\bf k}),
\end{equation}
where $N_0$ is the total number of particles in the Fermi gas. Similarly, the quasi-energy $\epsilon ({\bf k})$ also becomes a $2 \times 2$ matrix such that the change in the total energy, $\delta E$ due to a change in the distribution function $\delta n$ is given by:
\begin{equation}
    \delta E = V \int \frac{d^3 k}{ (2 \pi)^3} \epsilon_{\alpha \beta} ({\bf k}) \delta n_{\beta \alpha} ({\bf k}).
\end{equation}
We note that if the quasi-particle distribution function and the energy do not exhibit any spin dependence then, $n_{\alpha \beta} = n \delta_{\alpha\beta}$ and $\epsilon_{\alpha \beta} = \epsilon \delta_{\alpha\beta}$. Finally we find that  the change in the quasi-particle energy $\delta \epsilon_{\alpha \beta} ({\bf k})$ due to a change in the distribution function, $\delta n ({\bf k})$ is:
\begin{equation}
 \delta \epsilon_{\alpha \beta} ({\bf k}) = \sum_{\gamma, \delta} \int \frac{d^3 k^{\prime}}{ (2 \pi)^3} f_{\alpha \gamma, \, \beta \delta} ({\bf k}, {\bf k^{\prime}}) \delta n_{\delta \gamma} ({\bf k}),
\end{equation}
where 
\begin{equation}
  f_{\alpha \gamma, \, \beta \delta} ({\bf k}, {\bf k^{\prime}})  =  f_{\gamma \alpha , \, \delta \beta } ({\bf k^{\prime}},{\bf k})
\end{equation}
The spin-dependence of $f$ primarily arises from the exchange interaction, such that
\begin{equation}
 f_{\alpha \gamma, \, \beta \delta} ({\bf k}, {\bf k^{\prime}})  =  F ({\bf k}, {\bf k^{\prime}})\delta_{\alpha \beta} \delta_{\gamma \delta} + G ({\bf k}, {\bf k^{\prime}})\sum_a \sigma^{a}_{{\alpha \beta}}.\sigma^{a}_{\gamma \delta}, 
 \label{eq:FLsu2}
\end{equation}
where $\sigma^{a}$ are the Pauli spin matrices. We note that on the Fermi surface, $F$ and $G$ only depend on the angle $\theta$ between ${\bf k}$ and ${\bf k^{\prime}}$. The expression in eq.~\eqref{eq:FLsu2} originates from the independence of the exchange interaction from the spatial orientation of the total angular momentum. This ensures that the two spin operators appear as a scalar product. \\

This theory was generalized to the SU($N$) scenario by Yip, Huang, and Kao~\cite{yip2014theory} (see Ref.~\cite{cheng2017n} and Ref.~\cite{capponi2016phases} for extensions to finite temperatures and one-dimension respectively). They did this by describing the change in the quasi-particle energy at ${\bf k}$, $\delta \epsilon_{\alpha \beta} ({\bf k})$ as a $N \times N$ matrix, that is related to the low-energy quasi-particle excitations, $\delta n_{\delta \gamma }$ by:

\begin{equation}
\delta \epsilon_{\alpha \beta} ({\bf k}) = \sum_{{\bf k^{\prime}}, \gamma, \delta} f_{\alpha \gamma, \beta \delta} ({\bf k}, {\bf k^{\prime}}) \delta n_{\delta \gamma} ({\bf k^{\prime}}),
\end{equation}
where 
\begin{equation}
f_{\alpha \gamma, \beta \delta} ({\bf k}, {\bf k^{\prime}})  = f_s  ({\bf k}, {\bf k^{\prime}}) \delta_{\alpha \beta} \delta_{\gamma \delta} + 4 f_m ({\bf k}, {\bf k^{\prime}}) \sum_a T^{a}_{\alpha \beta} T^{a}_{\gamma \delta},
\end{equation}
where the matrices $T^{a}$ are the generators of the SU($N$) group. It is now fairly straightforward to extend the results for Landau FL theory for SU($N$) fermions. There are some remarkable consequences of this generalization. For instance, the Stoner instability in the SU(2) FL leads to a continuous phase transition from the paramagnetic to the Ferromagnetic phase. However, for $N>2$, this transition can become discontinuous~\cite{pera2023itinerant,huang2023itinerant,pera2024beyond}. Another important quantity that is strongly influenced by $N$ is the isothermal compressibility, $\kappa = \frac{1}{n^2} \frac{d n}{d \mu}$ given by
\begin{equation}
  \frac{\kappa_0}{\kappa}
   = 1 + (N -1) \frac{2 k_F a}{\pi} \left[ 1 + \frac{ 2 k_F a} { 15 \pi} (22 - 4 \ln 2 ) \right].
\end{equation}
where $\kappa_{0}$ is the compressibility when $a=0$, $n$ is the particle density, and $\mu$ is the chemical potential. Thus, in the weakly interacting regime ($k_F a \ll 1$), the SU$(N)$ gas is effectively $(N-1)$-fold more repulsive than the SU(2)-gas. \\

\begin{figure}[]
\centering
 % left lower right upper
 \includegraphics[width=0.9\textwidth]{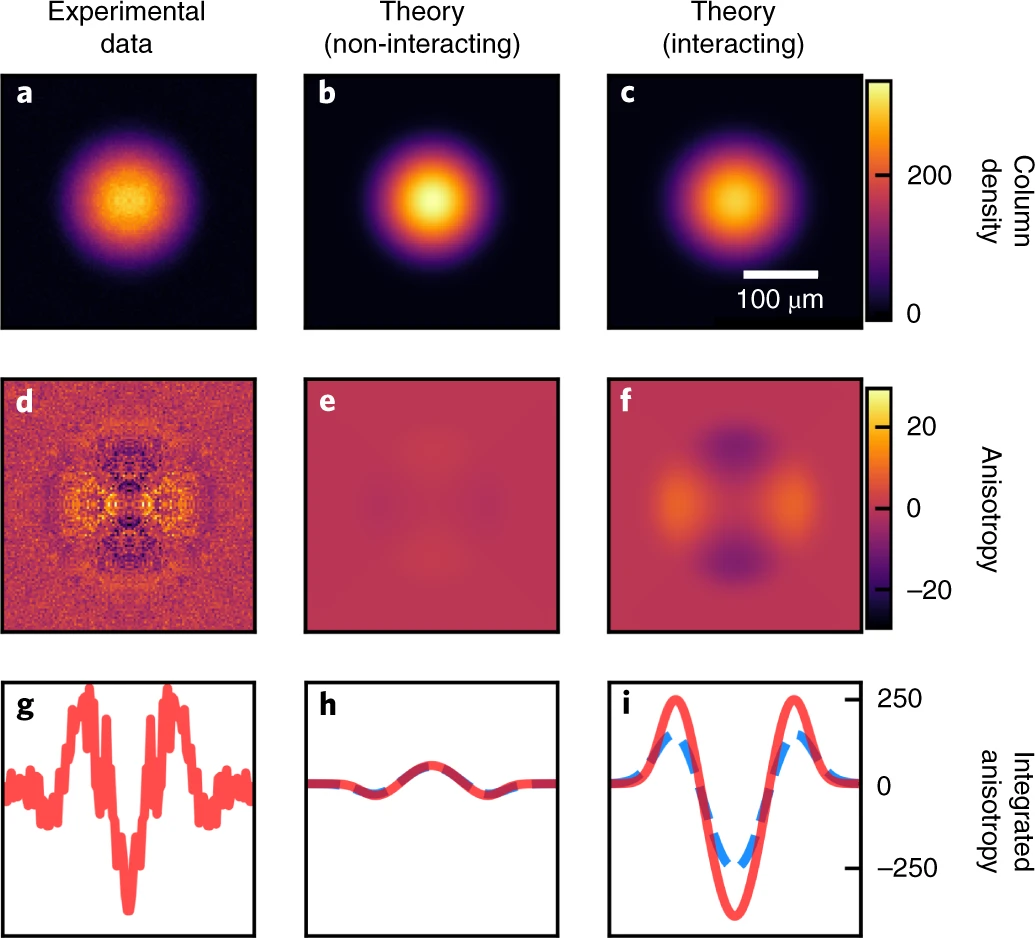}
\caption[Signatures of interactions in the degenerate SU($N$) Fermi Gas]{(a)-(c) shows the integrated atomic density obtained after time-of-flight. The anisotropy of the cloud shown in (d)-(f) reflects the anisotropy of the trap. (g)-(i) shows the anisotropy integrated along one direction. The red lines shows the anisotropy of the images in (d)-(f), while the dashed blue lines correspond to the anisotropy obtained using a higher temperature. Figure reprinted with copyright permission of Ref.~\cite{sonderhouse2020thermodynamics}.}
\label{fig::Sonderhouse_2020}
\end{figure}

Sonderhouse {\it et al.} measured this compressibility experimentally by examining the local density fluctuations~\cite{sonderhouse2020thermodynamics}. Employing the fluctuation-dissipation theorem, the relative number fluctuations, $\eta$ of a small sub-region of the gas having an average of $\overline{N_s}$ atoms can be related to $\kappa$ as $\eta = \Delta \overline{N_s}^2/\overline{N_s} = n k_BT \kappa$, where $k_B$ is the Boltzmann constant.  To the first order in both temperature and the scattering length, $a$, one obtains:
\begin{equation}
\eta = \frac{3}{2}\frac{T/T_F}{1+ \frac{2}{\pi} (k_F a)(N-1)},
\end{equation}
which clearly shows the enhanced effect of the repulsive interaction. We note that this expression can also be obtained by a virial expansion of the partition function. From their measurements, the authors find that $T/T_F = 0.16 \pm 0.01$, thereby indicating that the gas is in the deeply degenerate regime. However, it is interesting to note that the effect of interactions on the density fluctuations can also be mimicked by lowering the temperature of a non-interacting gas. Thus, to better characterize the interaction effects, the experimentalists also investigated the dynamics of the gas after being released from the trap. In the expansion dynamics, the interactions lead to a preferential movement of the atoms in the direction of the largest density gradient lead to an anisotropic distribution of the cloud after sufficiently long-times. The experimental results are shown in Fig.~\ref{fig::Sonderhouse_2020}; the strong role of interactions is evident here.\\

\subsection{Bosonization and Collective Modes}
One of the most intriguing features of a $N-$component Fermi gas is that this system would exhibit bosonic behavior in the large-$N$ limit. This bosonization arises from a weakened impact of the Pauli exclusion principle due to the large number of internal states and is well-established in one-dimensional systems. In 2012, the Florence group demonstrated this one-dimensional bosonization in Ref.~\cite{pagano2014one}. Intriguingly, recent theoretical results and experimental investigations have provided strong evidence for bosonization in higher dimensions. This has been achieved by probing the $N$-dependence of the collective mode frequencies and the contact in these systems. We now proceed to describe these studies.\\

\begin{figure}[htbp!]
\centering
 % left lower right upper
 \includegraphics[width=0.97\textwidth]{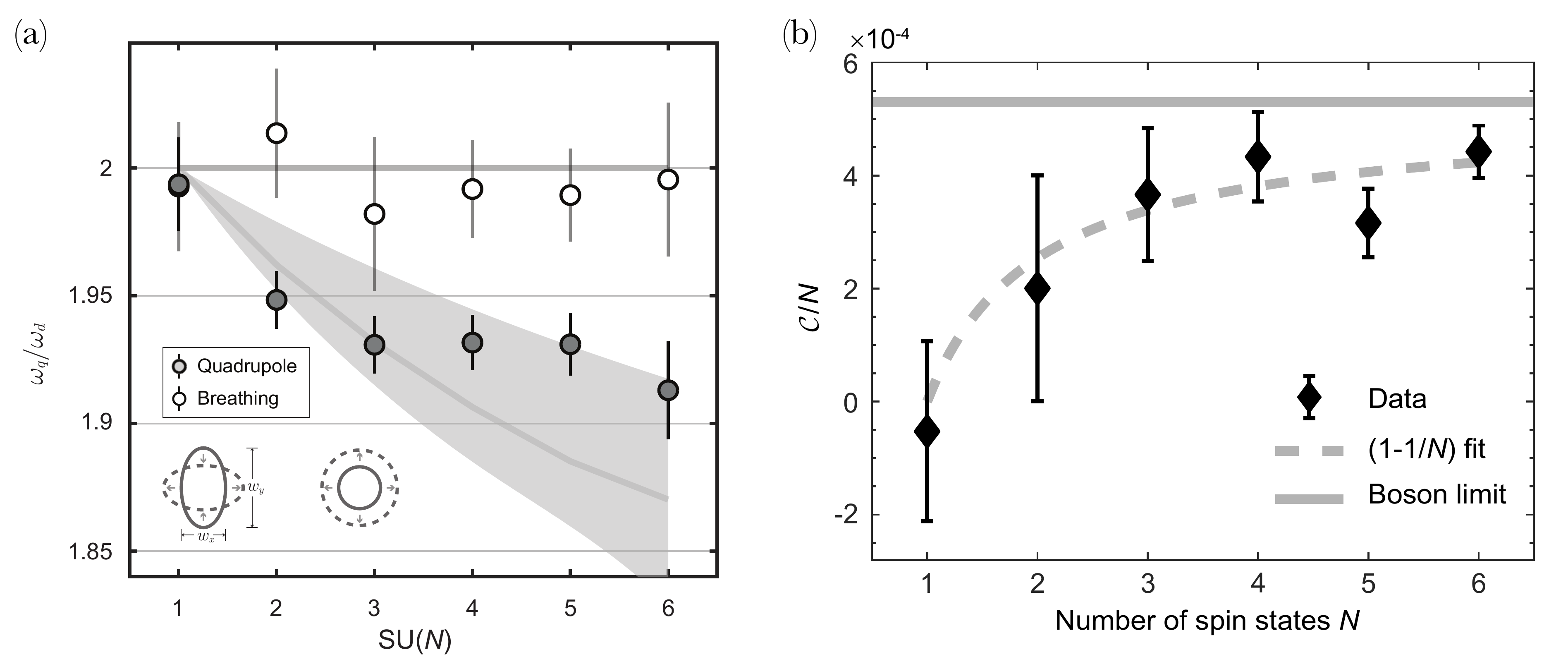}
\caption[Collective mode frequencies and Bosonization of the SU($N$) Fermi Gas]{(a) The $N-$dependence of the breathing and quadrupole mode frequencies of a trapped gas of $^{173}Yb$ atoms in two dimensions. The breathing mode does not show any $N-$dependence due to a classical scale invariance, while the quadrupole modes show a strong $N-$dependence. Figure reprinted and modified with permission from Ref.~\cite{he2020collective}. (b) The s-wave contact $C$ of SU($N$) fermions shows a strong $N-$dependence approaching the bosonic value, $C_B$ as $N$ increases ($C_{{\rm SU}(N)} \sim C_B (1-1/N)$). This shows the weakened effect of the Pauli exclusion principle.  Figure reprinted and modified with permission from Ref.~\cite{song2020evidence}.}
\label{fig::Bosonization}
\end{figure}

The collective oscillations of trapped quantum gases in response to external perturbations provide important insights into the many-body physics of these systems. These oscillations have been extensively studied in the case of two-component Fermi gases, where the collective mode frequencies and their damping rates reveal the effects of interactions. He {\it et al.} have extended these studies to the case of AEA by examining the breathing and quadrupole modes of a two-dimensional Fermi gas of $^{173}$Yb; they performed the experiment by suddenly  increasing the radial trap frequency, thereby exciting multiple collective modes~\cite{he2020collective}. The frequency of each collective mode was separately extracted by tracking the center-of-mass and the cloud width. The results obtained are shown in Fig.~\ref{fig::Bosonization}(a). It is clear that the breathing mode does not show any dependence on the spin multiplicity and stays at the value of $2 \omega_d$, where $\omega_d$ is the dipole mode frequency; this is a consequence of classical scale invariance in a weakly interacting two-dimensional gas. In contrast, mean-field effects lead to a clear dependence of the quadrupole mode frequency, $\omega_q \propto \left(2 \omega_d - g_{2D}(N-1)\right)$. We note that these experiments were performed in the `collisionless' regime. In contrast, in the hydrodynamic regime, collisions become extremely important. In this case, the quadrupole-mode frequency is $N-$independent $\omega_q^{\rm hd} = \sqrt[2] \omega_d$. Finally, we note that the damping rate of these oscillations, $\frac{1}{\tau} \propto (N-1)$, thereby highlighting the enhanced effect of the interactions~\cite{choudhury2020collective}.  \\

While collective modes play an important role in characterizing the many-body physics of Fermi gases, the thermodynamic properties of a dilute quantum gas can be captured succinctly by the contact, $C$~\cite{tan2008energetics,tan2008generalized,tan2008large}. In particular, $C$ governs various thermodynamic quantities via universal relations. In Ref.~\cite{song2020evidence}, Song {\it et al.} measured the $N-$dependence of s-wave contact $C$ from the momentum distribution of the Fermi gas. Due to the spin-independent nature of the interactions, the large-momentum tail scales as $n({\bf k}) = C_0/|{\bf k}|^4$. where $C_{{\rm SU}(N)} = c_{\rm pair} N_{\rm tot}^2 (1-1/N)$, where $N_{\rm tot}$ is the total number of fermions. Interestingly,  the large-momentum tail for spinless bosons is given by $C_B = c_{\rm pair} N_{\rm tot} (N_{\rm tot}-1) \approx c_{\rm pair} N_{\rm tot}^2$. Thus $C_{{\rm SU}(N)}$ approaches $C_B$ with a $1/N$ scaling thereby demonstrating bosonization in three dimensions.

\section{Quantum simulation with ultracold AEAs in optical lattices}
\label{sec:OL}

Quantum simulation with ultracold atoms in optical lattices (OLs) has provided with an unparalleled avenue to study many-body Hamiltonians relevant to condensed matter physics~\cite{Bloch2012,Gross2017,Schafer2020,Altman2021}. One the primary directions of the field is the experimental study of the Fermi-Hubbard model (FHM)~\cite{Hubbard_original,Montorsi1992,Tasaki1998,Arovas2022}. The FHM model is central to condensed matter physics since it is one of the simplest models that captures the essential features of strongly correlated materials and because it accounts for many canonical correlated phases of matter these systems exhibit. 
For example, in the two-dimensional (2D) square lattice, it displays many of the phenomena observed in strongly correlated materials such as the Mott insulating phase, long-range antiferromagnetic order, charge density waves, strange metallicity, a pseudogap, spin-charge ``stripe'' domains, and $d$-wave pairing~\cite{Imada1998,White1989,Schafer2021,Qin2020,Qin2022,Bohrdtreview,Zheng2017,Bourgund2023}. \\

In its original SU(2) symmetric form, the FHM describes the dynamics of spin-$1/2$ particles on a lattice with a nearest-neighbor tunneling amplitude $t$, and an on-site interaction $U$:
\begin{equation}\label{eq::FHM}
    H = -t \sum_{\sigma = \uparrow, \downarrow} \left[ \sum_{\langle i,j \rangle} \left( c_{i \sigma}^\dagger c_{j \sigma}^{\phantom{\dagger}} 
+ \mathrm{h.c.} \right) - \mu \sum_{i} n_{i \sigma} \right] + U \sum_{i} n_{i \uparrow} n_{i \downarrow}, 
\end{equation}
where $\expect{i,j}$ denotes nearest neighbors, $c_{i \sigma}^\dagger$ ($c_{i \sigma}^{\phantom{\dagger}} $) is the creation (annihilation) operator for a fermion with spin $\sigma$ on-site $i$, $n_{i \sigma} = c_{i \sigma}^\dagger c_{i \sigma}^{\phantom{\dagger}}$ is the number operator for spin $\sigma$ on site $i$, and $\mu$ is the chemical potential that controls the fermion density. The FHM successfully captures the physics of alkali atoms loaded in OLs. In these experiments, different hyperfine states of the alkali atom are used to represent the possible spin projections $\pm 1/2$. The density is set by controlling the number of particles $N_\uparrow$ and $N_\downarrow$ loaded into the lattice. The tunneling rate $t$ is controlled by changing the lattice depth, and the interaction strength $U$ is tuned via a magnetic Feshbach resonance. For further details, Refs.~\cite{Bloch2012,Gross2017,Schafer2020} provide a comprehensive review of the capabilities of these quantum simulators and their experimental tools.\\

\begin{figure}[htbp!]
\centering
\includegraphics[width=0.58\linewidth,angle=0]{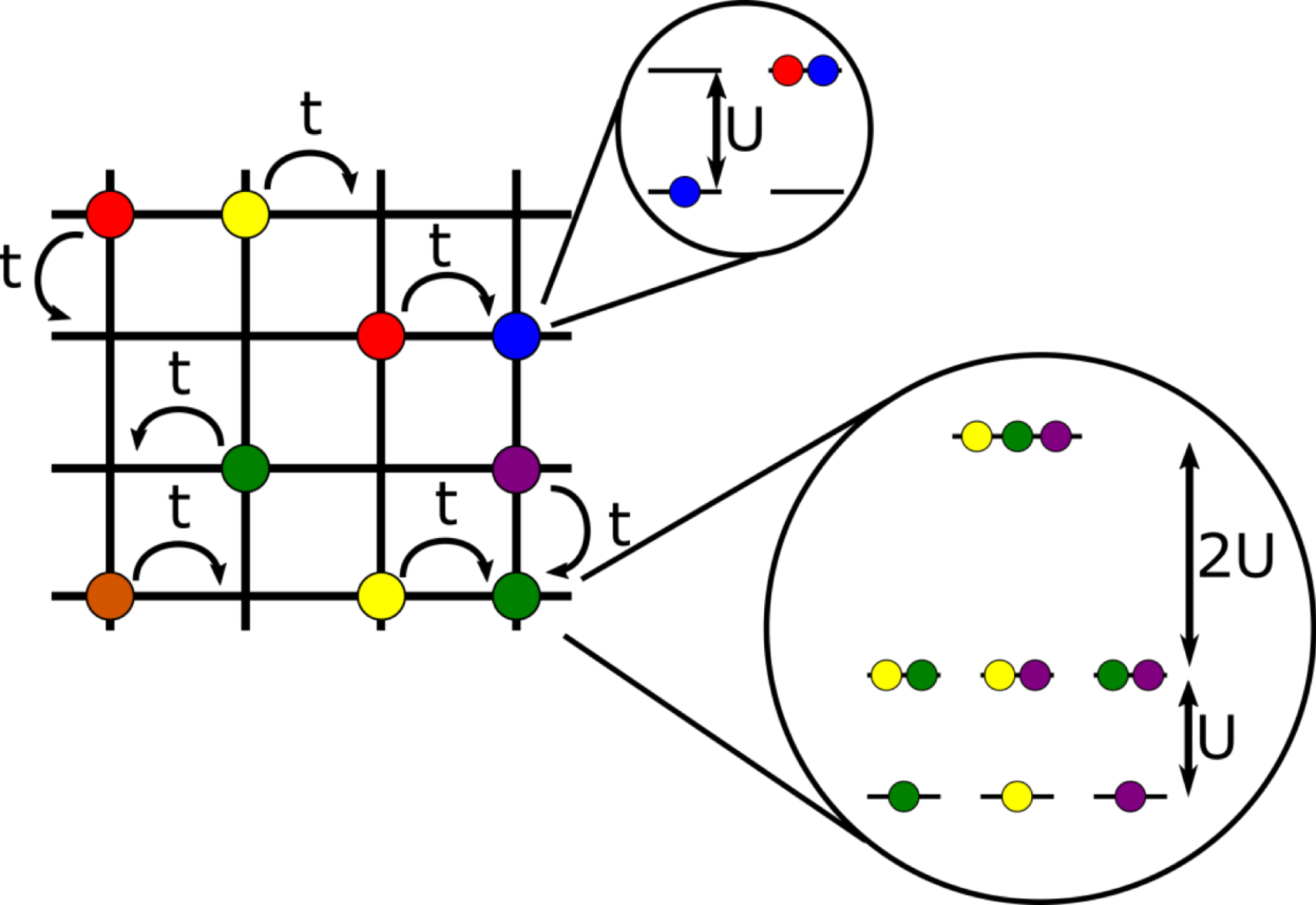}
\caption{In the SU($N$) FHM, particles with spin flavor $\sigma =1,\dots, N$ (denoted by different colors) live on a lattice. These particles can tunnel to neighboring sites with a hopping amplitude $t$. When $m$ atoms of different color occupy the same site, the energy is raised by $Um$. The filling fraction is controlled by a uniform chemical potential $\mu$ (not depicted here). Figure reprinted with copyright permission of Ref.~\cite{IbarraGarciaPadilla_thesis}.}
\label{fig::SUN_FHM} 
\end{figure}

In the case of fermionic alkaline-earth-like atoms (AEAs) in their ground state, by selective populating nuclear spin projection states $m_I$ and loading them into an OL, experiments can engineer the SU($N$) FHM with tunable $N$ from $2,\dots,10$. The SU($N$) Hamiltonian is,
\begin{equation}\label{eq::SUN_FHM}
    H = -t \sum_\sigma \left[ \sum_{\langle i,j \rangle} \left( c_{i \sigma}^\dagger c_{j \sigma}^{\phantom{\dagger}} 
+ \mathrm{h.c.} \right)  - \mu \sum_{i} n_{i \sigma} \right] + \frac{U}2 \sum_{i,\sigma \neq \tau} n_{i \sigma} n_{i \tau}, 
\end{equation}
where $\sigma =1,\dots,N$ are the spin flavors, and $N=2I+1$, where $I$ is the nuclear spin of the atoms. The Hamiltonian is graphically depicted in Fig.~\ref{fig::SUN_FHM}~\footnote{The interest in SU($N$) symmetric Hamiltonians is not limited to Hubbard and Heisenberg models, but also in two-band models such as the SU($N$) Kondo Lattice Model (KLM)~\cite{Coqblin1969,Doniach1977} which is commonly used in the study of manganese oxide perovskites~\cite{Tokurabook} and heavy-fermion materials~\cite{Coleman2007}. However, as of today, experimental efforts have mostly studied the single-band SU($N$) FHM, and we will therefore focus our attention on this model in this review. Further discussion in how the SU($N$) KLM can be engineered with AEAs in OLs is presented in Refs.~\cite{Gorshkov2010,Takahashi_review}.}. Analogous to the SU(2) FHM, the Hamiltonian in Eq.~\eref{eq::SUN_FHM} exhibits a U(1) symmetry which is reflective of a global charge conservation, i.e. the total number of particles is conserved, $[H,\sum_{i\sigma}n_{i\sigma}] = 0$.  In addition, the Hamiltonian is SU($N$) symmetric, since the energy scales $t,U,\mu$ are independent from $\sigma$. Experimentally, this means that (1) since $t$ and $\mu$ are controlled in a similar fashion to the alkali case, balanced mixtures of $N_\sigma$ atoms can be loaded into a lattice with a desired depth which is independent of the spin flavor. (2) The spin independence of the interaction term, i.e. $U_{\sigma,\tau} = U \, \forall \, \sigma,\tau$ arises from the fact that fermionic AEAs feature an almost perfect decoupling of the nuclear spin $I$ from the electronic structure in the ground state. Because of this decoupling, the s-wave scattering lengths $a$ for different $m_I$ exhibit predicted variations of the order of $10^{-9}$~\cite{Gorshkov2010,Cazalilla2014,Stellmer2014}, and therefore the interaction strength $U$ (which is proportional to $a$) is independent of $\sigma$.\\

Mathematically, the SU($N$) symmetry of Eq.~\eref{eq::SUN_FHM} means that the generators of the group, which are linear combinations of the spin permutation operators $S_\sigma^\tau = \sum_i c_{i \tau}^\dagger c_{j \sigma}^{\phantom{\dagger}}$, satisfy the SU($N$) algebra $[S_\sigma^\tau,S_\alpha^\beta] =  \delta_{\alpha \tau} S_\sigma^\beta - \delta_{\sigma \beta} S_\alpha^\tau$,
and commute with the Hamiltonian $[H,S_\sigma^\tau] \, \forall \, \sigma,\tau$. The SU($N$) symmetry is reflective of the spin isotropy, and similarly to the SU(2) case, the individual spin populations are conserved, since $S_\sigma^\sigma = \sum_i n_{i \sigma}$ commutes with $H$. For further details on SU($N$) group theory, ref.~\cite{Cazalilla2014} provides a useful brief digest, and ref.~\cite{Zee_Nutshell} is a comprehensive resource for group theory in general.  \\

Before we proceed to discuss the study of the SU($N$) FHM using AEAs in OLs, a few important remarks are important to discuss:
\begin{enumerate}
    \pagebreak
    \item \textbf{Quantum fluctuations are imporant in the SU($N$) FHM}. 
    
    \hspace{0.7cm} A common question we have encountered arises in terms of the $1/N$ expansion, and the assumption is that since the spin has been enlarged, the system should behave more classically. However, that is not the case. 

    \hspace{0.7cm} The $1/N$ expansion technique was first used to understand spin-$1/2$ systems that exhibit SU(2) spin symmetry. In this expansion, one reduces the role of quantum fluctuations by considering the classical limit of magnets with large spin $S$. In doing so, the expectation value of the spins acquires a definite value with a small variation around this saddle point and thus the $1/N$ expansion provides a method to therefore obtain mean field theories~\cite{Read1983,Affleck1985,Affleck1988,Bickers1987,Auerbach2012}. 

    \hspace{0.7cm} In the $1/N$ expansion, the relevant operators are the raising and lowering operators $S^{\pm}$, which only connect two possible values of the spin projections $\pm 1/2$, and therefore as $S$ increases, the variance of the spin projection falls off as $1/S$. In contrast, for the SU$(N)$ case even though $1/N$ can be small, the spin permutation operators $S_\sigma^\tau$ connect all possible values of the spin projection and therefore quantum fluctuations are relevant and play a major role in the ground state spin structure (see Fig.~\ref{fig::SUN_fluctuations}). For this reason, in the SU($N$) case the variance does not go like $1/N$.

    \begin{figure}[htbp!]
    \centering
    \includegraphics[width=0.68\linewidth,angle=0]{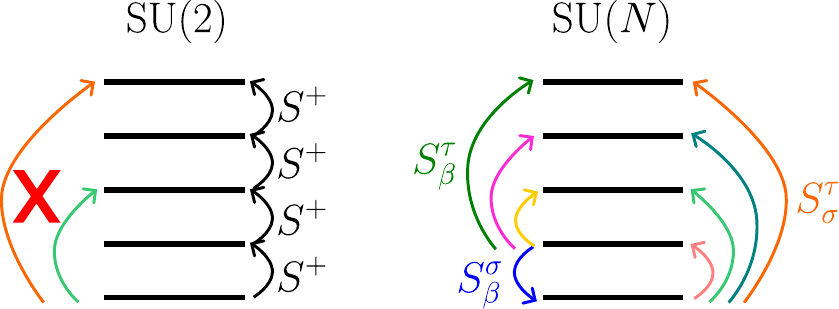}
    \caption{In the $1/N$ expansion, the underlying algebra is SU(2), and different states can only be reached by subsequent applications of the raising and lowering operators ($S^+$ depicted in the image). On the contrary for the SU($N$) algebra, the spin permutation operators $S_{\sigma}^\tau$ connect all possible states. While in the former one the variance of the spin expectation value decreases with the number of states, in the latter one quantum fluctuations play an important role in the physics.}
    \label{fig::SUN_fluctuations} 
    \end{figure}

\vspace{1cm}
    \item \textbf{The SU($N$) FHM is a limit of multi-orbitals models.}

    \hspace{0.7cm} Another common question is related to the connection to solid state systems. In particular, that if electrons are spin-$1/2$ particles, what do the spin flavors mean in the SU($N$) FHM?

    \hspace{0.7cm} In our introduction of the SU($N$) FHM we mentioned we have $N$ spin flavors, which in the context of quantum simulation with AEAs correspond to the nuclear spin projections $m_I$. In order to make the connection with the problem of electrons in a solid, we need to recall that an electron that is bound to or nearly localized on an specific site has three attributes: charge, spin, and orbital. While the SU(2) FHM considers a single orbital and the spin degree of freedom which is invariant under SU(2) rotations, multi-orbital models correspond to models in which higher orbitals are now considered. 
    
    \hspace{0.7cm} The SU($N$) FHM can be viewed as a limit of multi-orbitals models, such as the ones used to describe transition metal oxides~\cite{Li1998,Tokura2000,Dagotto2001}, in which the Hubbard parameters are independent of the spin and orbital degrees of freedom. It is worth noting that such enhanced symmetry is a crude approximation in solid state systems, where interactions between different orbitals may vary over 10\% or more~\cite{Tokura2000}. On the other hand, for ultracold atomic experiments the SU($N$) symmetry holds down to many orders of magnitude as previously discussed.

    \hspace{0.7cm} We have to make an important distinction regarding the origin of the enlarged symmetry in solid state systems and in quantum simulators with AEAs. While in the former ones the enhanced symmetry arises from the (degenerate) orbital and spin degrees of freedom, such that a pseudo-spin operator can be constructed to satisfy the SU($N$) algebra, in the latter ones it arises purely from the spin degree of freedom.  

\end{enumerate}
\vspace{0.5cm}

We now proceed to discuss the ground-breaking experimental achievements in quantum simulators using ultracold $^{173}$Yb OLs. For organizational purposes we separate these into two categories: (1) measurements of the equations of state, in which characterization of the SU($N$) FHM thermodynamic variables is explored and (2) quantum magnetism, in which the emphasis is focused on measuring spin correlation functions. We note that even without interactions, the spin states of AEA atoms can be employed to realize synthetic dimensions. In particular, the Florence group has realized synthetic hall ribbons with $^{173}$Yb atoms and observed topological edge states~\cite{Mancini2015}. However, in this work, we primarily focus on studies, where interactions play a crucial role.

\subsection{Equations of state of SU(N) FHM}
In this sub-section, we review some of the key experiments that investigated the equations of state of the SU($N$) FHM.
\subsubsection{SU(6) Mott Insulator}
\phantom{ }

In a pioneering experiment in 2012, the Kyoto group successfully achieved the realization of an SU(6)-symmetric Mott insulator in a three-dimensional (3D) optical lattice~\cite{Taie2012}. The importance of this study is three-fold: Firstly, this provided an experimental realization of an SU($N>2$) FHM. Secondly, it provided experimental evidence of a robust Mott plateau and the opening of charge gap via the demonstration of a suppression in the isothermal compressibility for $N=6$. Thirdly, it demonstrated a process analogous to ``Pomeranchuk'' cooling in solid $^3$He. \\

In this experiment, the SU($N$) FHM was realized by loading a balanced mixture of all the possible nuclear projections of $^{173}$Yb in its ground state in a three dimensional cubic optical lattice. The experiment started with a balanced mixture of all nuclear spin states which is evaporatively cooled in a crossed far-off resonant optical trap. At the end of evaporation, the experimentalists had $1-3 \times 10^4$ atoms with a temperature $ \sim 0.2 T_F$. Atoms were subsequently loaded into an optical lattice with lattice constant $d = 266$ nm. The lattice depth was varied from 6-13 $E_R$, to achieve $U/6t \in [1.8, 18.6]$~\footnote{The recoil energy $E_R$ is the natural energy scale for ultracold atom experiments and is defined as $E_R= \hbar^2 k^2/2m$. $k = 2\pi/\lambda$, $\lambda$ is the wavelength of the laser used to generate the optical lattice, and $d = \lambda/2$ is the lattice spacing.}.\\

\begin{figure}[htbp!]
\centering
 % left lower right upper
 \includegraphics[width=0.5\textwidth]{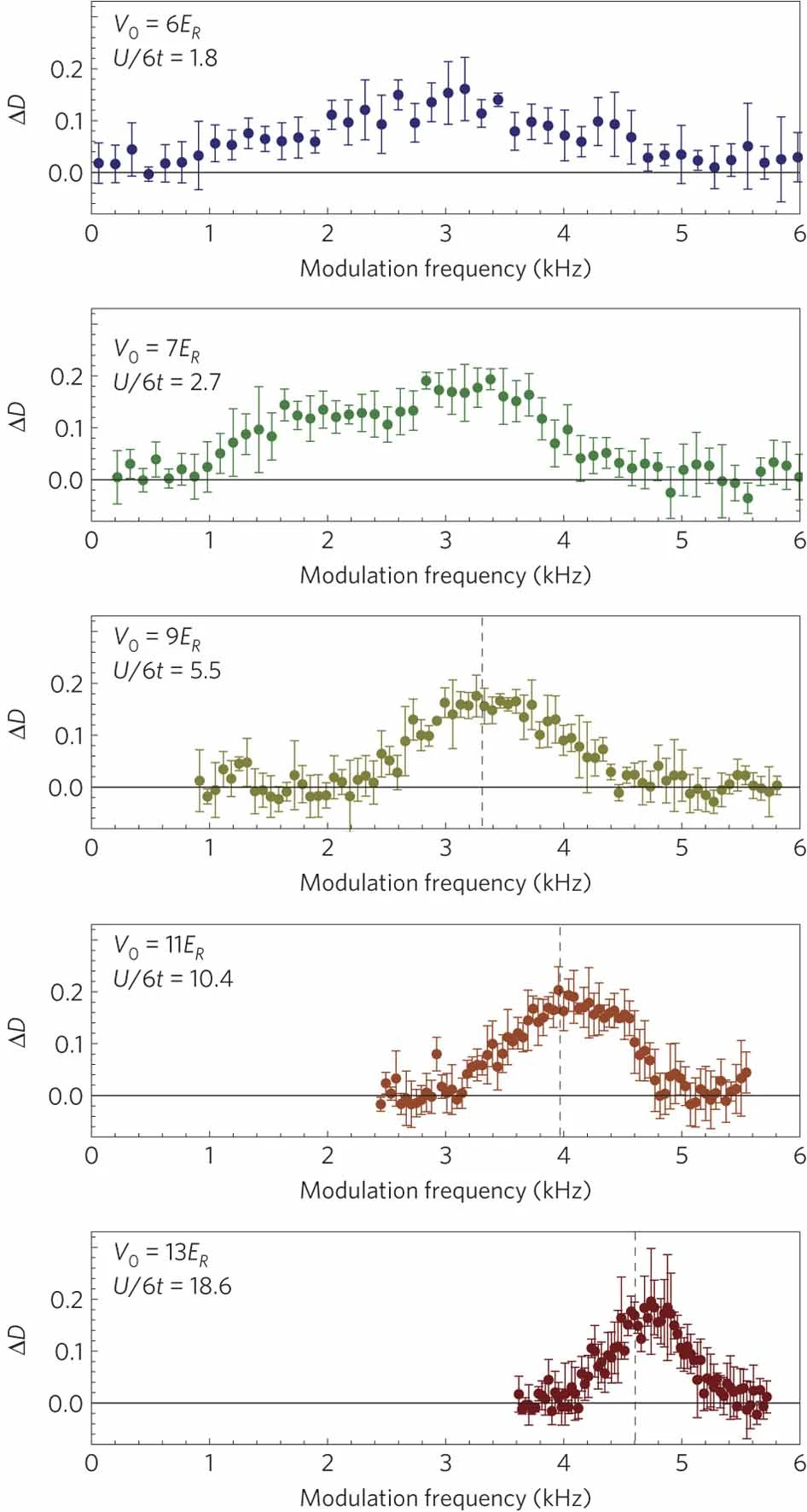}
\caption[Lattice modulation spectra]{ Lattice modulation spectra obtained for samples with $N = 1.9 \, (1) \times 10^4$ particles and initial entropy per particle $s/k_B = 1.9 \,(2)$, for different values of the lattice depth in $E_R$. The panels show the emergence of a peak centered around the frequency corresponding to the Mott gap (i.e. the measured values of on-site interaction $U$ for the corresponding lattice depth). Figure reprinted with copyright permission of Ref.~\cite{Taie2012}.}
\label{fig::Taie2012_1}
\end{figure}

\begin{figure}[htbp!]
\centering
 % left lower right upper
 \includegraphics[width=\textwidth,trim={0 0 0 0},clip]{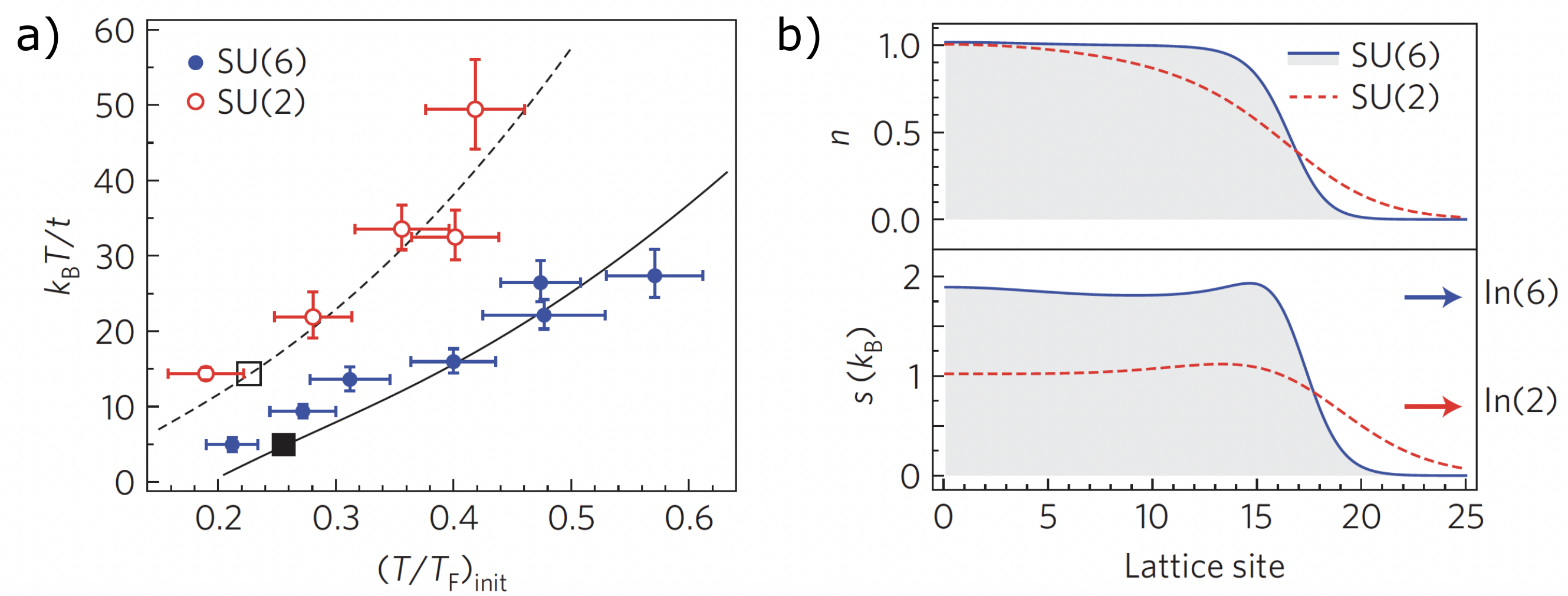}
\caption[SU($N$) Mott Insulators]{Results are presented for $N = 1.9 \, (1) \times 10^4$ particles and $U/6t = 10.47$.  a) Temperatures of the SU(6) (blue circles) and SU(2) (red open circles) Fermi gases after adiabatic loading in the lattice as a function of the initial temperature. b) Calculated density and entropy profiles at the lowest temperatures indicated by squares in a) for SU(6) (blue solid line) and SU(2) (red dashed line) using a second order high-temperature series expansion (HTSE). Figure modified with copyright permission of Ref.~\cite{Taie2012}.}
\label{fig::Taie2012_2}
\end{figure}

Figs.~\ref{fig::Taie2012_1} and~\ref{fig::Taie2012_2} summarize the main results of this experiment. In Fig.~\ref{fig::Taie2012_1}, the authors present experimental evidence of the charge gap in the SU(6) Mott insulator by performing lattice modulation spectroscopy~\cite{Jordens2008}. In this technique, the system is subjected to a periodic modulation of the lattice potential. If the modulation frequency $\nu$ roughly corresponds to the energy $\Delta E$ required to generate an excitation, that is, $h \nu \approx \Delta E$, then the system would, on average, absorb energy. In the case of the FHM, excitations between particles and holes are created at an energy of the order of the interaction strength $U$. Consequently, periodically modulating the lattice causes energy absorption when the frequency matches the excitation energy, that is, $h \nu \approx U$~\cite{Kollath2006}.
Therefore, by periodically modulating the lattice depth, resonant tunneling to the occupied sites at the modulation frequency close to the Mott gap $U$ is obtained and double occupancies are induced. 

The induced double occupancies can then be measured using photoassociation spectroscopy~\cite{Rom2004,Sugawa2011,photoassociation_review,photoassociation_review2}. ``Photoassociation is the process in which two colliding atoms absorb a photon to form an excited molecule''~\cite{photoassociation_review}. Consequently, all atoms on doubly occupied sites can be transformed into electronically excited molecules by the photoassociation process, which rapidly escape from the trap. Therefore, the loss of atom is the measure of the double occupancy. The Mott gaps are clearly observed in Fig.~\ref{fig::Taie2012_1} at higher lattice depths.\\

We now discuss the final important achievement of this experiment - the demonstration of ``Pomeranchuk" cooling with SU($N$) fermions. Historically, Pomeranchuk cooling was first predicted in the context of $^{3}$He~\cite{pomeranchuk1950theory}, where solidifying liquid $^{3}$He at low temperatures ($< 0.3 K$) leads to cooling~\cite{richardson1997pomeranchuk}. This intriguing effect originates from the higher entropy associated with the symmetry-broken solid phase compared to the itinerant Fermi liquid. In the Fermi liquid phase, only fermions near the Fermi surface contribute to the entropy at low temperatures; this contribution scales linearly with the temperature, $T$. In contrast, in the localized solid phase, each site with spin-$1/2$ contributes an entropy of $\ln(2)$ as long as the temperature $T$, and the spin exchange coupling strength, $J$ are of the same order of magnitude. When a mixture of solid and liquid $^{3}$He is compressed adiabatically at low temperatures, the liquid partially solidifies and the solid has the same entropy as the liquid that it replaces. This adiabatic redistribution of the entropy leads to cooling.\\

In the context of the Kyoto group experiment~\cite{Taie2012}, ``Pomeranchuk" cooling refers to the process where the entropy from the motional degrees of freedom of the Fermi gas is transferred to the spin degrees of freedom, thereby resulting in a lowering of the temperature of the system. The experimental evidence of ``Pomeranchuk'' cooling for the $N=6$ Fermi gas is presented in Fig.~\ref{fig::Taie2012_2}, which was originally predicted in refs.~\cite{Assaraf1999,Hazzard2012}. This cooling is reflected in the final temperature of the sample, and it was observed by comparing results for $N=2$ and $N=6$, for which the final temperature in the lattice after adiabatic loading (i.e. total number of particles and entropy are conserved) is lower for $N=6$ than for $N=2$ (see Fig.~\ref{fig::Taie2012_2}b). The lower temperature for larger $N$ is a consequence of entropy contribution of an isolated spin, which is $\propto \ln(N)$. Thus, for large system sizes, the Mott insulating state can be cooled to much lower temperatures, when $N$ is large. In addition, Fig.~\ref{fig::Taie2012_2}c demonstrates that for the lowest temperature $T/t$ achieved in each case, only the $N=6$ case develops a robust Mott plateau at the center of the trap with an entropy per site close to $\ln(N)$.

\subsubsection{Mott Crossover in three-dimensional OLs}
\phantom{ }

The Munich group has investigated the equation of state (EoS) of the SU(3) and SU(6) FHMs in a cubic 3D optical lattice~\cite{Hofrichter2016}. The relevance of this study is three-fold: First, the experimental investigation of the EoS for the density $n(\mu,T,N,U)$ for a wide range of chemical potentials and interaction strengths. Second, the experimental evidence of the metal-to-insulator crossover via the experimental determination of the local compressibility. Third, the lack of thermometry in the experiments at interaction strengths of the order of the bandwidth ($U \approx W$) served as a motivation for the development of more sophisticated numerical methods to analyze this system. \\

In this experiment, the SU($N$) FHM was realized by preparing a degenerate Fermi gas of $^{173}$Yb with $N=6$ equally populated spin components via evaporative cooling in a crossed dipole trap. At the end of evaporation, the Munich group had $5\times10^3$ atoms per spin state at temperature $T=0.07 T_F$. For the experiments with $N=3$, they remove individual spin components by driving the $^1S_0 \to$ $^3P_1$ optical transition in the presence of a homogeneous magnetic field that lifts the spin-state degeneracy, and are left with an SU(3) Fermi gas at $T=0.15 T_F$ with a residual fraction of unwanted spin components below 5\%. Atoms are then loaded into an optical lattice with lattice constant $d = 380$ nm. The lattice depth was varied from 3-15$E_R$ to achieve $U/12t \in [0.128,11.0]$.\\

\begin{figure}[htbp!]
	\centering
		\includegraphics[width=\textwidth]{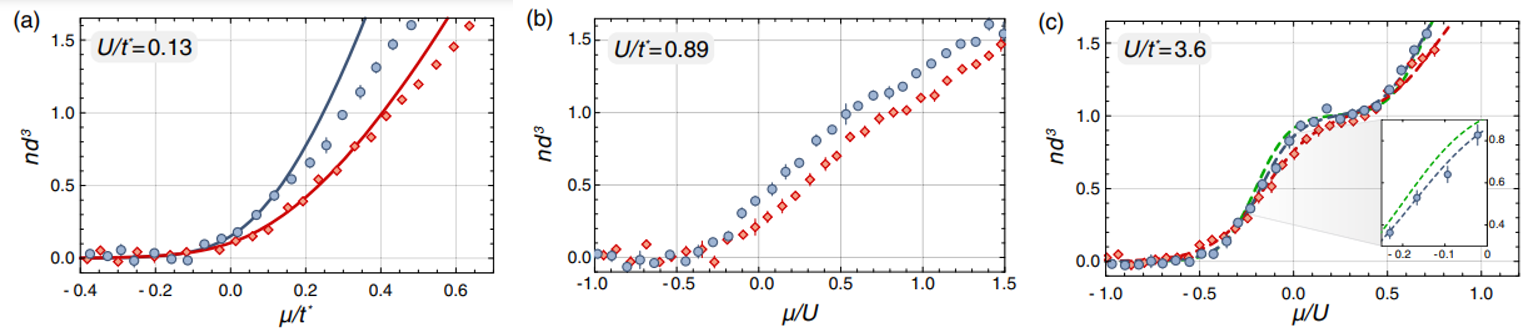}
	\caption[Equation of State of the SU($N$) FHM]{Density as a function of the chemical potential for $N=3$ (red diamonds) and $N=6$ (blue squares) Fermi gases in a 3D lattice. Here $t^* = 12 t = W$ is the non-interacting bandwidth of the 3D lattice. a) $U/t^*=0.128$ b) $U/t^*=0.89$ c) $U/t^*=3.6$. Solid lines are fits to the non-interacting Fermi gas EoS for densities below $0.5$. Dashed lines are a second-order high temperature series expansion to extract the temperature (in green for $N=2$ for comparison). Figure modified with copyright permission of Ref.~\cite{Hofrichter2016}.}
	\label{fig::Hofrichter2016_1}
    \end{figure}

\begin{figure}[htbp!]
    \centering
	\includegraphics[width=0.85\textwidth]{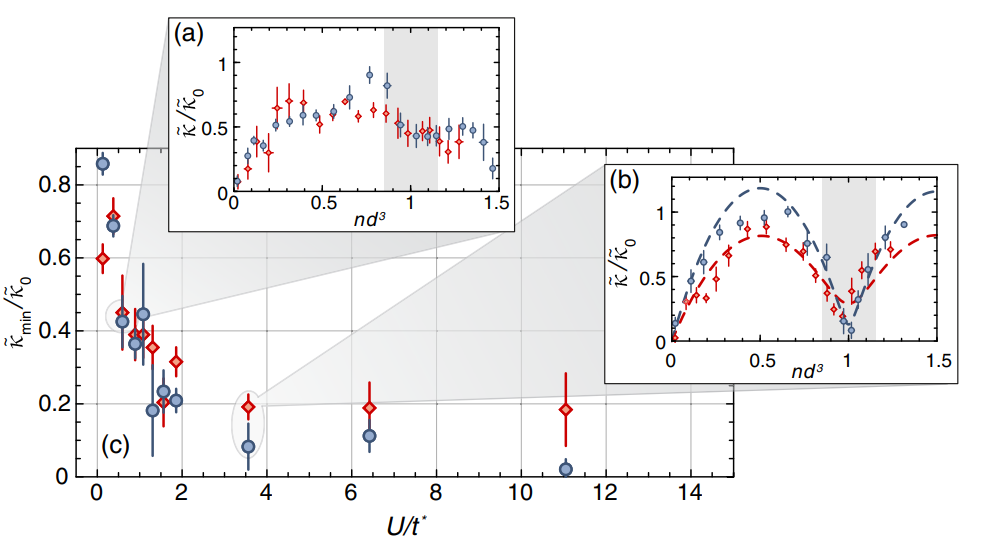}
    \caption[Compressibility of the SU($N$) FHM]{Compressibility for $N=3$ (red diamonds) and $N=6$ (blue squares) Fermi gases in a 3D lattice. Here $t^* = 12 t = W$ is the non-interacting bandwidth of the 3D lattice. Insets correspond to the compressibility as a function of density at a) $U/t^*=0.59$, and b) $U/t^*=3.6$. Dashed lines are a second-order high temperature series expansion. c) Minimal compressibility $\tilde{\kappa}_\mathrm{min}$ as a function of the interaction strength, where $\tilde{\kappa}_0$ is the compresssibility of the non-interacting SU(6) FHM at $\expect{n}=1$. Figure reprinted with copyright permission of Ref.~\cite{Hofrichter2016}.}
	\label{fig::Hofrichter2016_2}
    \end{figure}

Figs.~\ref{fig::Hofrichter2016_1} and~\ref{fig::Hofrichter2016_2} summarize the main results of the experiment. In Fig.~\ref{fig::Hofrichter2016_1} the authors present the experimentally measured density as a function of the chemical potential for different values of the interaction strength $U/t$. These results highlight that for $U \ll t$, the system is metallic and can be approximately described by the non-interacting theory for $n <1$ (see Fig.~\ref{fig::Hofrichter2016_1}a), whereas for $U \gg t$, the single site limit is a good approximation and can provide a good interpretation of the data (see Fig.~\ref{fig::Hofrichter2016_1}c). However, for interactions of the order of the bandwidth $U \sim W$ (Fig.~\ref{fig::Hofrichter2016_1}b), the system is a strongly correlated many-body state and at the time of publication in 2016 there were no numerical techniques to compare against.\\   

In Fig.~\ref{fig::Hofrichter2016_2}, the authors exploit the model-free access to the EoS for the density to measure the local compressibility $\tilde{\kappa} = n^2 \kappa = \partial \expect{n} / \partial \mu \vert_T$. In the strongly-interacting case [see Fig.~\ref{fig::Hofrichter2016_2}(b)], the compressibility is suppressed around $\expect{n}=1$, which is consistent with the opening of charge gap and the development of a Mott insulator. In addition, the Mott crossover is studied in Fig.~\ref{fig::Hofrichter2016_2}(c) where the minimum of the compressibility in the region $0.85<\expect{n}<1.15$ is presented as a function of the interaction strength. The minimum in the compressibility exhibits a suppression of roughly 1 order of magnitude and saturates at a minimum value for large $U/t$, indicating the system is deep in the Mott insulating state.

\subsubsection{EoS and Mott Crossover in two-dimensional OLs}
\phantom{}

Building upon their previous work, the Munich group recently realized a precise characterization of the equation of state of the SU($N$) FHM in the two-dimensional (2D) square lattice for $N=3, 4, 6$, and the results of their experiment are reported in Ref.~\cite{Pasqualetti2024}. The relevance of this study lies in: (1) The implementation of a 2D single-layer SU($N$) ensemble which can be probed with perpendicular absorption imaging with a resolution of a
few lattice sites. (2) The opportunity to benchmark state-of-the-art numerical methods that were recently developed and adapted to explore the SU($N$) FHM in experimentally accessible regimes, such as determinant quantum Monte Carlo (DQMC) and Numerical Linked Cluster Expansions (NLCE)~\cite{IbarraGarciaPadilla_thesis,IbarraGarciaPadilla2021,IbarraGarciaPadilla2023}. (3) The experimental demonstration of thermometry for SU($N$) Fermi gases in OLs in a model-independent way using the fluctuation-dissipation theorem (FDT).\\

In this experiment, the SU($N$) FHM was realized by first loading a spin-balanced mixture of approximately $1.6 \times 10^6$ $^{173}$Yb atoms from a magneto-optical trap into a crossed optical dipole trap, where evaporative cooling is performed down to $T/T_F^{(3D)} <0.2$. Subsequently, a second stage of evaporative cooling is performed in the presence of an optical gradient, yielding an ensemble of $N\sim 2\times10^3$ atoms in the central plane of a vertical lattice with wavelength $\lambda = 759$ nm and lattice spacing $d_\mathrm{vertical} = 3.9 \mu$m. In this configuration, the authors implemented a 2D single-layer SU($N$) ensemble with lattice spacing for $d=380$ nm. The implementation of a single-layer is crucial, as it avoids integration over inhomogeneous stacks of 2D systems. It also allows for direct access to the density profile without the reconstruction techniques previously required in Ref.~\cite{Hofrichter2016}, and which allows for the measurement of density fluctuations. The density distribution is measured using \textit{in situ}, saturated absorption imaging with a spatial resolution of approximately $2 \mu \mathrm{m} \approx 5d$. \\

    \begin{figure}[htbp!]
	\centering
		\includegraphics[width=\textwidth,trim={0 4.85cm 0 0},clip]{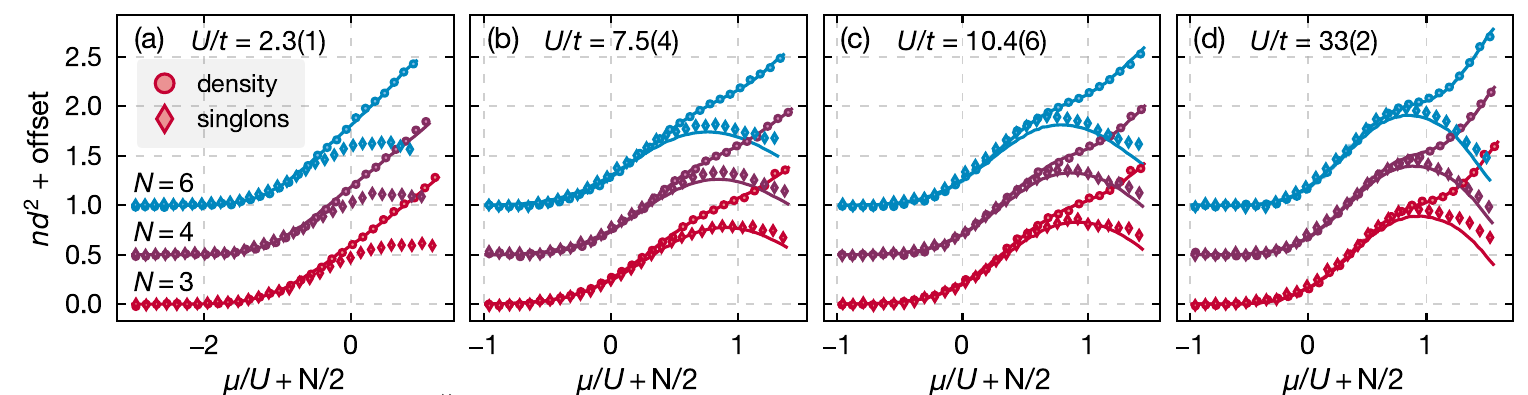}
	\caption[EoS 2D SUN FHM]{Equation of state for the SU($N$) FHM for $N=6$ (blue), $N=4$ (purple), and $N=3$ (red). Density (circles) and parity projected measurements (diamonds) are presented as a function of the chemical potential for different interaction strengths (a) $U/t=2.3$, (b) $U/t = 7.5$, (c) $U/t=10.48$, and (d) $U/t =33.2$. Solid lines associated to the density curves correspond to the fit of the EoS to DQMC (a) and NLCE (b)-(d) to realize thermometry. The results from the fit models are also used to calculate the parity projected measurements. All spin mixtures were prepared with the same initial entropy per particle $s/k_B = 1.2$ in the bulk before loading into the lattice. Figure modified with copyright permission of Ref.~\cite{Pasqualetti2024}.}
	\label{fig::Pasqualetti2}
    \end{figure}

   \begin{figure}[htbp!]
	\centering
		\includegraphics[width=0.8\textwidth,trim={0 0 0 8cm},clip]{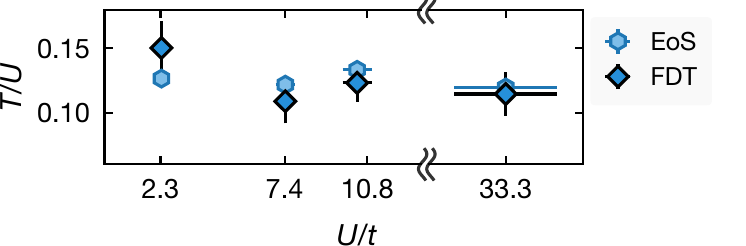}
	\caption[bbb]{Comparison of the temperatures obtaind using the fluctuation-dissipation theorem (FDT) [dark blue diamonds] and the fits to the equation of state (EoS) [light blue hexagons]. Error bars are the standard error of the mean. Figure modified with copyright permission of Ref.~\cite{Pasqualetti2024}.}
	\label{fig::Pasqualetti3}
    \end{figure}

The main results of this experiment are summarized in Figs.~\ref{fig::Pasqualetti2} and~\ref{fig::Pasqualetti3}. In Fig.~\ref{fig::Pasqualetti2} the authors present the experimentally measured densities (circles) and photoassociated parity projected measurements (diamonds) as a function of the chemical potential for different values of $N$ and the interaction strength $U/t$. The fit of the EoS was performed in two dimensions, with the temperature $T$ and the chemical potential at the center of the trap $\mu_0$ as the free parameters, where the local density approximation was used to incorporate the contributions of the trap confinement~\cite{Nascimbene2010}. For $U/t = 7.5$ and $U/t =10$ experiments fitted both DQMC and NLCE, observing an excellent agreement between the theory and the experiment yielding consistent fitting parameters for the two different numerical methods. For $U/t=33$, results from NLCE and a second order HTSE also display excellent agreement. For $U/t=2.3$ the temperature lies below the range of convergence of NLCE and experiments resorted to DQMC alone.\\

In addition to the total density, in Fig.~\ref{fig::Pasqualetti2} the authors also characterized the number of double occupancies in the model by removing doublons via photoassociation. When available, the NLCE prediction (lines) based on the density's fit and without any further free fit parameters was compared to the experimental observations (diamonds), and they correspond well with the experimental data. Because of the possibility to directly access the density in the 2D single-layer, measurement of density fluctuations could be performed. By measuring the density fluctuations, the FDT connects the variance of the detected atom number in an area $A\gg d^2$ to the isothermal compressibility $\kappa$ and the temperature $T$ via
\begin{equation}
    \kappa A = \frac{1}{k_B T} \mathrm{var} \left( \int_A n dA \right),
\end{equation}
providing model-free thermometry~\cite{Qi2011,Hartke2020}. In Fig.~\ref{fig::Pasqualetti3}, the authors compared the temperature obtained using the FDT (squares) against the temperature returned by the fit to the EoS (hexagons), and observed good agreement for all interaction strengths.

\subsubsection{Flavor-selective localization in three-dimensional OLs}
\phantom{}

\begin{figure}[htbp!]
	\centering
\includegraphics[width=0.8\textwidth]{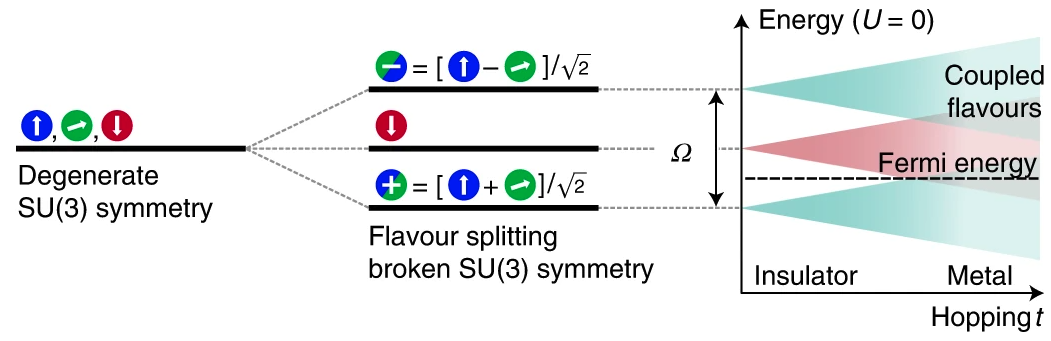}
	\caption[bbb]{The Raman coupling lifts the degeneracy between the states, creating two dressed states with different energies. The competition with the hopping can drive a metal-to-insulator transition already in the non-interacting case. Figure modified with copyright permission of Ref.~\cite{Tusi2021}.}
	\label{fig::Tusi1}
    \end{figure}

In a recent experiment, the Florence group realized flavor-selective Mott localization in an SU(3) Fermi gas in a 3D optical lattice~\cite{Tusi2021}. This study is relevant because it corresponds to: (1) The experimental realization of multicomponent Hubbard physics with coherent internal couplings. (2) The experimental achievement of studying Mott physics while explicitly breaking the SU(3) symmetry. In this experiment, a three-component ultracold $^{173}$Yb mixture with total atom number $N = 4\times10^4$ at an initial temperature of $T\approx 0.2T_F$ was used, which was loaded into a 3D cubic OL with lattice constant $d= 380$ nm. As previously discussed, AEAs in OLs are well described by the SU($N$) symmetric FHM [see eq.~\eref{eq::SUN_FHM}]. In this experiment, the authors explicitly broke the SU($N$) symmetry by introducing the following term
\begin{equation}\label{eq::Raman}
    H_R = \frac{\Omega}{2} \sum_i \left( c_{i \sigma}^\dagger c_{i \tau}^{\phantom{\dagger}} 
+ \mathrm{h.c.} \right),
\end{equation}
which describes a coherent on-site coupling between spin flavors $\sigma$ and $\tau$. This coupling is provided by a two-photon Raman process with Rabi frequency $\Omega$. At the single particle level, this Raman coupling lifts the degeneracy between the spin flavors, by creating two dressed states $|\pm \rangle = (|\sigma \rangle \pm |\tau \rangle)/\sqrt{2}$, with energy shifts $\pm \Omega/2$ relative to the other spin flavors (see Fig.~\ref{fig::Tusi1}).\\

The coherent coupling between states $m_I = 5/2$ and $m_I=1/2$ was realized by using a two photon $\sigma^+/\sigma^-$ Raman transition. In this scheme, the Raman coupling is implemented with two co-propagating laser beams with wavelength $\lambda = 556$ nm and angular frequencies $\omega$ and $\omega +\delta \omega$, which are blue-detuned by $1.754$ GHz with respect to the $^1S_0 \to$ $^3P_1$ ($F=7/2$) recombination transition to reduce inelastic photon scattering. A 150 G magnetic field is used to define a quantization axis and to lift the degeneracy between the six hyperfine states of $^{173}$Yb ground state manifold, which are split by $207 \times m_I$ Hz/G. The $\sigma^+/\sigma^-$ coupling between $m_I = 5/2$ and $m_I=1/2$ is obtained by setting the polarization of the two beams to be orthogonal with respect to the quantization axis and by adjusting $\delta \omega/2\pi$ to compensate the Zeeman splitting and the residual Raman light shift between the two states~\cite{Mancini2015}. For measurements at $\Omega=0$, the authors utilized optical pumping of the $^1S_0 \to ^{3}P_1$ transition to prepare a balanced mixture in the hyperfine states $m_I = \pm 5/2, 1/2$. While for measurements at $\Omega \neq 0$, the loading procedure started with with a 2-component unbalanced mixture of atoms in states $m_I = \pm 5/2$, such that $N_{5/2} = 2N/3$ and $N_{-5/2} = N/3$. Then after loading the lattice, the Raman beams were turned on far detuned from any two photon transitions, and an adiabatic frequency sweep was performed to bring them resonant to the $5/2 \leftrightarrow 1/2$ transition. This procedure corresponds to an adiabatic passage that brings an atom in the $m_I = 5/2$ state to the lowest energy dressed state $|+ \rangle = (|5/2\rangle + |1/2 \rangle)/\sqrt{2}$. At the end of the process, the population in equally distributed between the three hyperfine states.\\
    
     \begin{figure}[htbp!]
	\centering	\includegraphics[width=\textwidth]{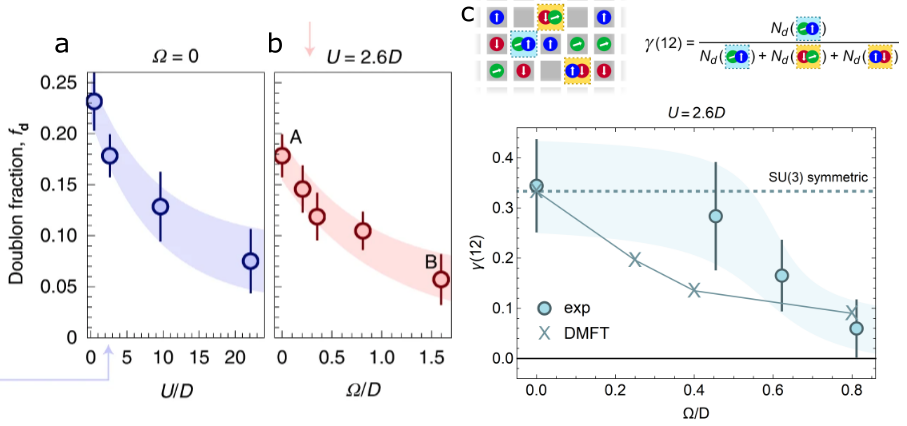}
	\caption[State-selective correlations]{Average doublon fraction $f_d$ as a function of  (a) $U/D$ at $\Omega =0$ (b) $\Omega/D$ at $U/D = 2.6$. Here $D = 6 t = W/2$ is half the bandwidth. (c) $\gamma(12)$ measures the number of atoms forming doublons in the $|\frac{5}{2},\frac{1}{2}\rangle$ channel, normalized by the total number of atoms forming doublons. Figure modified with copyright permission of Ref.~\cite{Tusi2021}.}
	\label{fig::Tusi4}
    \end{figure}

The main results of this experiment are summarized in Fig.~\ref{fig::Tusi4}. In Figs.~\ref{fig::Tusi4}(a)-(b) the authors study the behavior of the number of double occupancies as an indicator of the degree of the Mott insulating nature of the system. While in Fig.~\ref{fig::Tusi4}(a) the emergence of an SU(3) Mott insulator is indicated by the expected suppression of the doublon fraction as the interaction strength increases, Fig.~\ref{fig::Tusi4}(b) illustrates that the doublon fraction decreases as the Rabi coupling increases, leading to a similar Mott localized state.  Fig.~\ref{fig::Tusi4}(c) demonstrates that the double occupancies are flavor-dependent when $\Omega >0$. In this figure, $\gamma(12) = N_d(12)/N_d$ is the number of atoms forming doublons in the $|\frac{5}{2},\frac{1}{2}\rangle$ channel normalized by the total number of atoms forming doublons. In the absence of Raman couplings ($\Omega=0$) the results agree with the $N=3$ symmetric expectation value (dotted line). As $\Omega$ increases, the SU(3) symmetry is broken and $\gamma(12)$ diminishes, approaching zero as $\Omega \approx D = 6 t = W/2$ half the bandwidth. Doublons acquire a flavor-selective behavior, since doublon formation in the $|\frac{5}{2},\frac{1}{2}\rangle$ channel is suppressed because it requires fermions in both the $|\pm\rangle$ states which have an additional energy cost of $\Omega/2$ in contrast to the other two channels $|-\frac{5}{2},\frac{1}{2}\rangle$ and $|\frac{5}{2},-\frac{5}{2}\rangle$.

\subsection{Quantum Magnetism}
In this sub-section, we review some pioneering experiments that investigated quantum magnetism in the SU($N$) FHM.

\subsubsection{Antiferromagnetic spin correlations in a dimerized lattice}
\phantom{}

In Ref.~\cite{Ozawa2018}, the Kyoto group reported their measurements of nearest-neighbor antiferromagnetic (AFM) correlations in a Fermi gas with SU(4) symmetry in an optical superlattice. The importance of these results are two-fold. First, at a fixed entropy per particle, AFM nearest-neighbor correlations are enhanced for $N=4$ compared with $N=2$. Second, it was the first experimental utilization of the single-triplet oscillation technique~\cite{Greif2013} to measure nearest-neighbor spin correlation functions for $N>2$, and set up the basis to measure those in lattices with uniform tunnelings, which were further reported in Ref.~\cite{Taie2022}. In this experiment, the $N=2$ and the $N=4$ FHMs were realized by loading $^{173}$Yb in an optical superlattice with wavelengths $\lambda = 1064$ and $\lambda =532$ nm, for the long and short lattice, respectively. During evaporative cooling, optical pumping is performed to create SU(2) or SU(4) samples. After loading into the lattice, nearest-neighbor AFM correlations between adjecent sites in the optical superlattice were measured using the singlet-triplet oscillation (STO) technique, which we describe below.\\

\begin{figure}[htbp!]
	\centering
		\includegraphics[width=0.75\textwidth,trim={0 8cm 0 0},clip]{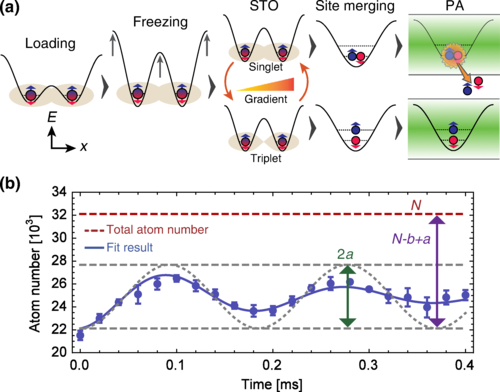}
	\caption[STO method]{(a) Detection sequence for singlets and triplets in a dimer. Shown is the case of two spins per dimer. Figure modified with copyright permission of Ref.~\cite{Ozawa2018}.}
	\label{fig::ozawa2018_sto}
    \end{figure}

The STO technique is presented in Fig.~\ref{fig::ozawa2018_sto}. In this method, after atoms are loaded into the lattice, the lattice depth is ramped up to suppress tunneling, and freeze atoms in place. Then a spin-dependent gradient beam is used to induce oscillations between the singlet and triplet states. Depending on the STO time, the spins form a doublon in the lowest band or a state with one spin in the lowest band and the other one in the first excited band after sites are merged. Performing photoassociation, the double occupancies are removed and the associated particle loss is measured, which is proportional to the spin-spin correlation function. \\

    \begin{figure}[htbp!]
	\centering
		\includegraphics[width=\textwidth]{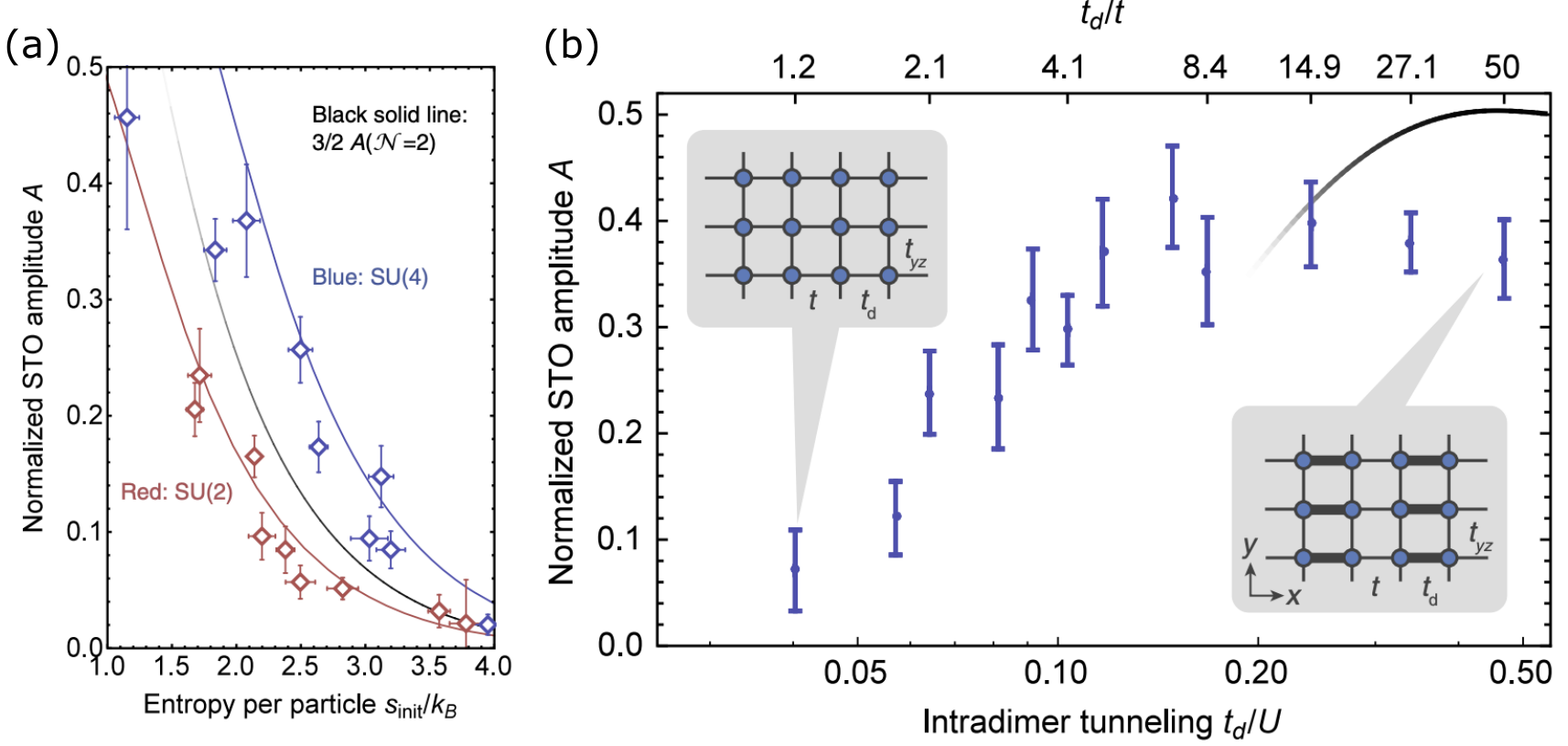}
	\caption[Nearest-neighbor AFM correlations in a dimerized lattice]{The normalized STO amplitude in a dimerized lattice as a function (a) the initial entropy per particle and (b) the intradimer tunneling $t_d$ at $U/h = 3.0$ kHz. (a) Red (blue) markers correspond to $N=2$ ($N=4$) and results are presented for $t_d/t =27$. (b) Results for $N=4$. Here the tunneling rates vary from $t \in [28.0,100]$ Hz, and $t_{yz}/t \in [1.7,1.0]$. Figure modified with copyright permission of Ref.~\cite{Ozawa2018}.}
	\label{fig::ozawa2018}
    \end{figure}

The main results are summarized in Fig.~\ref{fig::ozawa2018}, where the authors present the normalized STO amplitude $A$ as a function of (a) the initial entropy per particle and (b) the intradimer tunneling. In Fig.~\ref{fig::ozawa2018}(a) the authors show that the normalized STO amplitude decreases as the initial entropy per particle increases since the triplet states become thermally populated. Furthermore, they demonstrate that antiferromagnetic correlations are enhanced in the SU(4) system compared to SU(2) for the same initial entropy. This enhancement is due to the difference of the fraction of singlet configurations among all possible states $N \choose 2$, and the cooling effect related to spin entropy that was previously observed~\cite{Taie2012}. In Fig.~\ref{fig::ozawa2018}(b) the authors discuss the dependence of $A$ on the intradimer tunneling rate $t_d$. As $t_d$ decreases, the normalized STO amplitude decreases because the excitation energy to the triplet state is lowered, which in the two particle two site sector, is given by $-U/2 + \sqrt{16 t_d^2 +U^2}/2$. The experimental data suggests that although nearest-neighbor antiferromagnetic correlations get smaller as $t_d$ is lowered, they should retain a non-vanishing amplitude in the isotropic lattice.

\subsubsection{Antiferromagnetic spin correlations in OLs with uniform tunnelings}
\phantom{}

Building upon their previous work in strongly dimerized optical lattices, the Kyoto group then measured SU($N$) antiferromagnetic nearest-neighbor spin correlations in OLs with isotropic tunnelings. Besides being the first experimental determination of SU($N$) AFM correlations in lattices with homogeneous tunnelings, an important milestone of this study corresponds to the creation of the coldest fermions ever created in nature in absolute temperature and in cold atoms. In this experiment, spin-balanced mixtures of $2.4 \times 10^4$ $^{173}$Yb atoms are adiabatically loaded into 1D, 2D and 3D cubic OLs with lattice constant $d=266$ nm. To achieve the lower dimensional lattices, a strong tunneling anisotropy is introduced in one or two directions to suppress tunneling in that direction (the inter-lattice tunneling is $\lesssim 5\%$ than the intra-lattice tunneling). \\

After loading into the lattice, nearest-neighbor AFM correlations were measured using the STO technique. Similarly to Ref.~\cite{Ozawa2018}, the application of a spin-dependent potential gradient before the merging process drives oscillations between the singlet and the triplet states in two adjacent sites. Such spin-dependent potential gradient is generated by applying an optical Stern–Gerlach laser beam close to the $^1S_0 \to$ $^3P_1$ resonance, with a detuning of $+2.6$ GHz from the $F=5/2 \to 7/2$ transition, which minimizes the ratio of the photon scattering rate to the differential light shifts. Experiments measure the fraction of both singlet and triplet states formed within nearest-neighbour lattice sites. It is important to notice that the naming ``singlet'' and ``triplet'' in this context corresponds to the SU($N$) counterparts of the SU(2) double-well singlets and triplets, and should not be confused with SU($N$) singlets and triplets, which are $N$-body entangled states~\cite{Li1998}.\\

The detected SU($N$) counterpart of the SU(2) double-well singlet is a $N \choose 2$-fold multiplet of the form $(|\sigma,\tau \rangle - |\tau, \sigma \rangle)/ \sqrt{2}$, while the the double-well triplet is a [$N \choose 2$ + $N$]-fold multiplet, among which the $N \choose 2$ states of the form $(|\sigma,\tau \rangle + |\tau, \sigma \rangle)/ \sqrt{2}$ are detected by the STO scheme, while the case $\sigma=\tau$ is not. The STO measurement is valid only if the contribution from multiple occupancies can be neglected, and therefore the authors set the density at the center of the trap to one particle per site on average, and set the interaction strength to a large value of $U/t=15.3$ to strongly supress the formation of double and higher occupancies. \\

\begin{figure}[htbp!]
	\centering
		\includegraphics[width=0.5\textwidth,trim={0 0 8cm 0},clip]{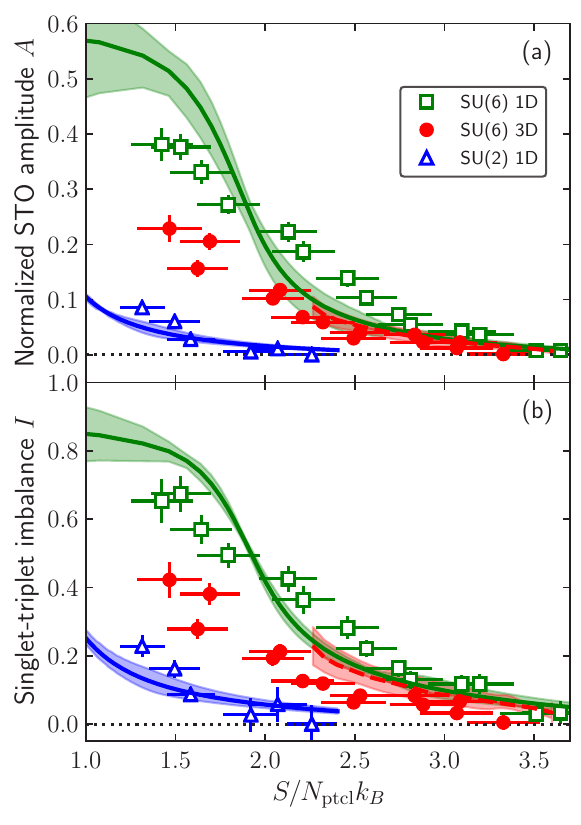}
	\caption[Nearest-neighbor AFM correlations] {Entropy dependence of the (a) normalized STO amplitude $A$ (b) singlet–triplet imbalance I. Results are presented for 1D and 3D lattices, showing experimental data for the SU(6) 1D (green squares), SU(6) 3D (red circles) and SU(2) 1D systems (blue triangles). Horizontal error bars represent the standard deviation of ten entropy measurements, while vertical error bars are extracted from the fitting errors in the analysis of the STO signal. Numerical calculations with ED (solid lines) and DQMC (dashed lines) are also displayed. Shaded areas represent uncertainty from the systematic and statistical errors of the numerical methods plus the possible systematic error ($20\%$) in the total atom number measurement. Figure modified with copyright permission of Ref.~\cite{Taie2022}.}
	\label{fig::taie2022_2}
    \end{figure}

    \begin{figure}[htbp!]
	\centering
		\includegraphics[width=\textwidth]{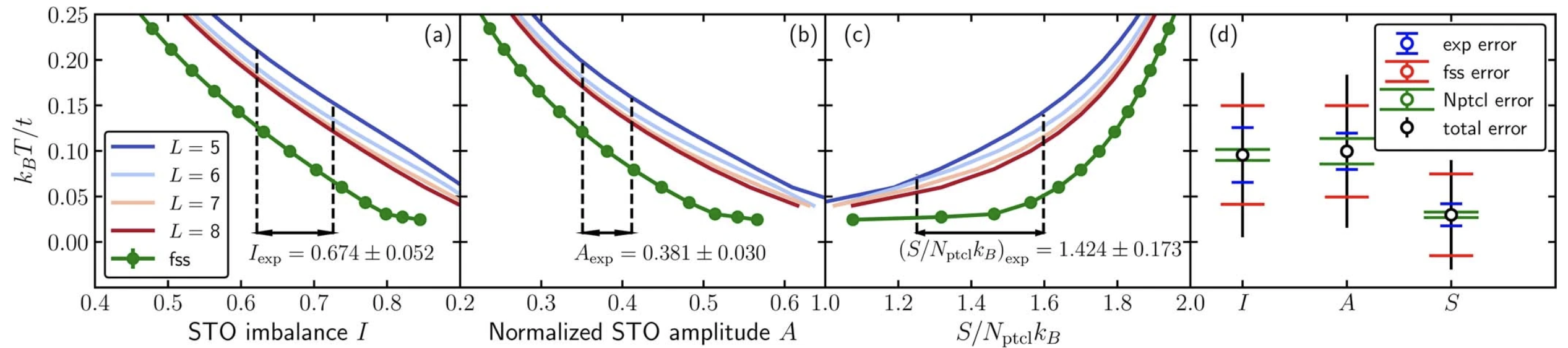}
	\caption[Thermometry SU6] {Thermometry of an SU(6) Fermi gas in a one-dimensional OL at $U/t=15.3$. (a)-(c) Solid lines correspond to exact diagonalization results in $L$ sites chains. Green circles and lines correspond to finite-size extrapolations. Dashed vertical lines correspond to the range of the largest  experimentally measured STO imbalance in 1D that is consistent with error bars. (d) Extracted temperatures from the STO imbalance, amplitude and the initial entropy per particle. Error bars come from the measured correlations error bars (blue), the finite-size error (red), the particle number fluctuation (green), and their sum (black). Figure reprinted with copyright permission of Ref.~\cite{Taie2022}.}
	\label{fig::taie2022_4}
    \end{figure}

The main results of the manuscript are summarized in Figs.~\ref{fig::taie2022_2} and ~\ref{fig::taie2022_4}. In Fig.~\ref{fig::taie2022_2} the authors present the experimentally measured STO imbalance $I$ and amplitude $A$ for SU(6) and SU(2) Fermi gases in 1D and 3D OLs as a function of entropy per particle. The imbalance and the amplitude are defined as,
\begin{align}
    A &= -\frac{1}{N_\mathrm{ptcl}} [C_\mathrm{NN}]_\mathrm{total} \\
    I &= \frac{2A}{A +[n(i) n(i+1)]_\mathrm{total}/N_\mathrm{ptcl}},
\end{align}
where  $\expect{n(i)n(i+1)}$ is the nearest-neighbor density–density correlation function, and $C_\mathrm{NN}$ is the nearest-neighbor spin-spin correlation function,
\begin{equation}
    C_\mathrm{NN} = \sum_{\sigma \neq \tau} \bigg[ \expect{n_\sigma(i)n_\sigma(i+1)} - \expect{n_\sigma(i)n_\tau(i+1)}\bigg],
\end{equation}
which measures the likelihood of having a different spin flavors on adjacent sites~\footnote{It is easy to see that in the limit $N=2$, $C_\mathrm{NN}$ reduces to the SU(2) nearest-neighbor $\expect{S_zS_z}_\mathrm{nn}$ correlation function.}.  In the local-density approximation (LDA),
\begin{equation}
    [\mathcal{O}]_\mathrm{total} = \int \frac{d^3r}{d^3} \expect{\mathcal{O}(\mu(\textbf{r},T)}
\end{equation}
for an arbitrary observable $\mathcal{O}$.\\

Fig.~\ref{fig::taie2022_2} demonstrates that at fixed entropy per particle, AFM nearest-neighbor spin correlations get enhanced as $N$ increases, and also that these correlations appear at higher entropy for larger $N$. This last point can be easily interpreted deep in the Mott regime at $1/N$ filling, in which the minimum entropy per site attainable before any correlations develop is $\ln(N)$. As the temperature is lowered, entropy is lowered as correlations develop. Furthermore, it illustrates the dimensionality dependence, with the 1D case exhibiting the largest correlations. This behaviour is similar to previous studies in an SU(2) system~\cite{Imriska2014,Greif2013,IbarraGarciaPadilla2020} and can be understood at sufficiently high temperatures. In this regime, correlations depend only on temperature not dimension, and decrease with increasing temperature. Additionally, as the dimensionality decreases, the bandwidth also decreases, and therefore, at fixed entropy, the temperature decreases. These arguments together imply that correlations decrease as the dimensionality is increased. The authors observed that the experimental data show reasonable agreement with theoretical predictions obtained by exact diagonalization (ED) for 1D and by determinant quantum Monte Carlo (DQMC) for 3D, without any fitting parameters. In the case of $N=6$ in 3D, the experiments reach entropies below the regime where the calculations converge, highlighting the importance of these experiments as quantum simulations, and providing motivation for the development and refinement of numerical techniques to mitigate the sign problem to reach lower temperatures~\cite{Feng2023,Zhang2019}.\\

At that time, this experiment did not have the capabilities to directly measure the temperature at the very low entropies studied. However, for the 1D systems, the temperature was inferred by comparing experiment and theory. In 1D, the lowest temperature achieved in the experiments corresponds to $k_BT/t=0.096 \pm 0.054 \pm 0.030$, obtained from the experimentally measured STO imbalance $I$ at $S/N_\mathrm{ptcl}k_B=1.45 \pm 0.05$. The first error bar corresponds to the finite-size error of the calculations, and the second one corresponds to the experimental errors in the correlation measurements. The details of the thermometry are presented in Fig.~\ref{fig::taie2022_4}, where estimates of the temperature using $A$ instead of $I$, yield consistent results.  For comparison, to obtain the same singlet–triplet imbalance, the $N=2$ system should be at $S/N_\mathrm{ptcl}k_B=0.499 \pm 0.136 \pm 0.120$. Since the state-of-the-art experiments of the SU(2) FHM with alkali atoms in OLs perform at an entropy per particle $\sim 1k_B$ ($k_BT/t=0.25 \pm 0.2$)~\cite{Mazurenko2017}, this suggests an experimental advantage for SU$(N)$ systems in obtaining highly correlated states in optical lattices.\\

The experimental studies of the thermodynamic and magnetic properties of the SU($N$) FHM using AEAs in optical lattices pave the way toward more direct quantum simulation of typical 2D models of interest in naturally occurring systems with SU($N>2$) representations such as transition metal oxides and orbitally selective Mott transitions. A fascinating example is the cerium volume collapse, which has long been debated as to whether the single-orbital Hubbard model ($N=2$) or the double-orbital Hubbard model ($N=4$)~\cite{johanssonAgTransitionCerium1974,allenKondoVolumeCollapse1982,lippAnomalousElasticProperties2017,heldCeriumVolumeCollapse2001} is the proper description. Although the SU($N$) symmetry is usually only roughly realized in condensed matter cases, quantum simulators with AEAs offer an almost exact realization of SU($N$), enabling the implementation of fully SU($N$)-symmetric and formerly purely theoretical models. More generally, AEAs quantum simulations of the SU($N$) FHM can shed light on the validity of the SU($N$) approximation in more realistic models, and it should even be possible to smoothly connect both regimes in a continuous way by controlled symmetry-breaking using techniques like optical state manipulation or state-dependent potentials~\cite{Gorshkov2010,Tusi2021,Yi_2008}.

Furthermore, in the 2D square lattice, a question of importance in the SU(2) FHM arises at half-filling and finite dopings around $\expect{n}=1$. One of the relevant aspects of quantum simulation of the SU($N$) FHM at $1/N$-filling ($\expect{n}=1$), is that at this filling, the $N>2$ FHM enables us with the possibility to untwine the role played by nesting and Mott physics because, unlike SU(2), a finite $U$ is needed to open a charge gap. While the SU(2) FHM at $1/2$-filling shows a van Hove singularity in the density of states and perfect nesting of the Fermi surface, the $N>2$ counterparts achieve $\expect{n}=1$ without these unique band structure features. This enables us to isolate the effect of interactions from band structure attributes without having to take into account next-nearest-neighbor tunneling amplitudes $t'$, which are difficult to control in experiments involving ultracold atoms.

\subsection{Overview of computational and theoretical works on SU(N) lattice models} 

\begin{table}
\mysize %scriptsize was fine before, now the caption does not fit %tiny is too small
\begin{center}
\caption{\small Short compendium of the theory of SU($N$) Lattice models.}
\begin{tabular}{@{}llllll}
\br
Model \,\, & $N$ \,\, & Geometry \,\, & $T \,\, $ & Methods \,\, & References \,\, \\
\mr
Hubbard & 2-4, 6 & Square & F & DQMC, NLCE, HTSE, ED &\cite{IbarraGarciaPadilla2021} \\
Hubbard & 3 & Square & F & DQMC, NLCE & \cite{IbarraGarciaPadilla2023} \\
Hubbard & 3 & Square & 0 & CPQMC & \cite{Feng2023} \\
Hubbard & $\geq2$ & Square & F & RG, MF & \cite{Honerkamp2004} \\
Hubbard & 2,4,6 & Square & 0 & Projector QMC &  \cite{Wang2014} \\
Hubbard & 6 & Square & 0 & Projector QMC & \cite{Wang2019} \\
Hubbard & 2,4,6 & Square & F & DQMC & \cite{Zhou2014}\\
Hubbard & 2-4 & Square & F & HTSE & \cite{Rajiv2022a} \\
Hubbard & 4 & Square & 0 & Projector QMC & \cite{Zhou2018,Xu2024}\\
Hubbard & 4 & Square & F & DMFT & \cite{Goubeva2017} \\
Hubbard & 3-10 & Square & 0 & SRMF & \cite{Chen2016} \\ 
Hubbard & 4 & Square & F & DMFT & \cite{Unukovych2021} \\
Hubbard & 6 & Square & F & FLT & \cite{Cazalilla2009} \\
Hubbard & 2-6 & Square & 0 & ED & \cite{Botzung2024b} \\
Hubbard & 6 & Square & F & Diagrammatic QMC & \cite{Kozik2024} \\
Hubbard & 2,4,6,8 & Square & 0 & large-$N$, Projector QMC & \cite{Assaad2005} \\
Hubbard & 2,4,6 & Square, Honeycomb & 0 & Projector QMC & \cite{Ouyang2021} \\ 
Hubbard & 2-6 & Square, Honeycomb, Triangular & 0 & ED & \cite{Botzung2024a} \\
Hubbard & 3 & Square, Cubic, Bethe & F & DMFT, VMC & \cite{Titvinidze2011} \\ 
Hubbard & 3 & Square, Cubic & F & DMFT &  \cite{Sotnikov2014} \\
Hubbard & 3 & Cubic & F & DMFT & \cite{Sotnikov2015} \\

Hubbard & 2-5 & Chain & 0 & DMRG, BA, Bos & \cite{Manmana2011} \\
Hubbard & 2-4,6,10 & Chain & F & SSE-QMC & \cite{Bonnes2012} \\
Hubbard & 2,3,4 & Chain & 0 & GFMC &\cite{Assaraf1999} \\ 
Hubbard & 4-14 & Chain & F & SSE-QMC & \cite{Xu2018} \\
Hubbard & 3 & Chain & 0 & DMRG &\cite{PerezRomero2021} \\
Hubbard & 2-5 & Chain & 0 &  DMRG & \cite{Buchta2007}\\
Hubbard & 4 & Chain & 0 & BA, DMRG & \cite{Yamashita1998}\\
Hubbard & 2-6 & Chain & F & DMRG & \cite{Mikkelsen2023} \\

Hubbard & 4 & Honeycomb & F & HTSE & \cite{Rajiv2022b} \\
Hubbard & 4,6 & Honeycomb & 0 & Projector QMC & \cite{Zhou2016}\\
Hubbard & 4,6 & Honeycomb & F & DQMC & \cite{Zhou2017}\\
Hubbard & 3 & Honeycomb & 0 & Projector QMC & \cite{Xu2023} \\
Hubbard & 3 & Honeycomb & 0 & iPEPS & \cite{Chung_PRB_2019} \\

Hubbard & $\geq 2$& Bipartite 2D & F & HTSE & \cite{Rajiv2022c} \\

Hubbard & 3 & Triangular & F & DMFT & \cite{Hafez2018,Hafez2019,Hafez2020}\\

Hubbard & 2,3 & Lieb & 0,F  & MF &  \cite{Nie2017} \\

Hubbard & 3 & Bethe & 0 & DMFT & \cite{delRe_PRA_2018} \\

Hubbard & 2,4,6 &  & F & HTSE & \cite{Hazzard2012} \\
Hubbard & 2-5 & & F & DMFT & \cite{Lee2018} \\
Hubbard & 2-6 & & F & DMFT & \cite{Yanatori2016}\\

\mr
Heisenberg & 3 & Square, Cubic & 0 & ED & \cite{Toth2010} \\
Heisenberg & 3 & Square, Triangular & 0 & DMRG, iPEPS & \cite{Bauer2012} \\
Heisenberg & 4 & Square & 0 & ED, iPEPS & \cite{Corboz2011} \\
Heisenberg & $\geq 2$ & Square & 0 & large-$N$ & \cite{Hermele2009} \\
Heisenberg & 5,8,10 & Square & 0 & ED & \cite{Nataf2014} \\
Heisenberg & $\geq2$ & Square & 0 & large-$N$ & \cite{Hermele2011} \\
Heisenberg & 2-5 & Chain & F & CTWLMC & \cite{Messio2012} \\
Heisenberg & 3,4 & Chain, Square, Triangular & F & HTSE, ED & \cite{Romen2020} \\
Heisenberg & $>2$ & Chain & 0 & CFT, ED,DMRG & \cite{Herviou2023} \\
Heisenberg & 3 & Honeycomb & 0 & ED, iPEPS, VMC & \cite{corboz2013competing} \\
Heisenberg & 3 & Honeycomb & 0 & DMRG & \cite{zhao2012plaquette} \\
Heisenberg & 3 & Honeycomb & 0 & LFWT & \cite{lee2012spontaneous} \\
Heisenberg & 4 & Honeycomb & 0 & MF, VMC & \cite{Natori2019} \\
Heisenberg & 4 & Honeycomb & 0 & ED, iPEPS, VMC & \cite{Corboz_PRX_2012} \\
Heisenberg & 6 & Honeycomb & 0 & ED, iPEPS, VMC & \cite{Corboz_PRB_2016} \\
Heisenberg & 3 & Triangular & 0,F & CMF, semiclassical MC & \cite{Yamamoto2020} \\
Heisenberg & 2-9 & Triangular & 0 & MF & \cite{Yao2021} \\
Heisenberg & 4 & Triangular & 0 & MF, LFWT, ED & \cite{delRe_PRR_2024} \\
Heisenberg & 3 & Kagome & 0 & ED, iPEPS & \cite{Corboz_PRB_2012}\\
\mr

${t-J}$ & 3 & Square & 0 & DMRG & \cite{Henning2023} \\
${t-J}$ & 4 & Chain & 0 & DMRG & \cite{He2024}\\
\mr 

SSH & 2-5 & Square & 0 & Projector QMC & \cite{Yu_PRL_2024} \\
\br
\end{tabular}\label{numericalwork}\\
\end{center}
 
\textit{Abbreviations} $F$: finite-temperature. MF: mean-field, MC: Monte Carlo, QMC: Quantum Monte Carlo, BA: Bethe-Ansatz, Bos: Bosonization, CFT: conformal field theory, CMF: cluster MF, CTWLMC: continuous-time world-line MC, CPQMC: constrained path QMC, DMFT: dynamical mean-field theory, DMRG: density matrix renormalization group, DQMC: determinant QMC, ED: exact diaonalization, FLT: Fermi liquid theory, GFMC: Green's function MC, HTSE: high-temperature series expansions, iPEPS: infinite projected entangled-pair states, LFWT: Linear-flavour-wave theory, NLCE: numerical linked cluster expansion, RG: renormalization group, VMC: variational MC, SRMF: slave rotor mean-field, SSE-QMC: quantum MC simulations within the stochastic series expansion (SSE), SSH: Su-Schrieffer-Heeger.

\end{table}
\normalsize

Historically, the study of SU($N$) quantum magnetism arose from the mathematical technique of large-$N$ expansions~\cite{Auerbach2012,Read1983,Affleck1985,Affleck1988,Bickers1987}. However, the possibility of exploiting the inherent SU($N$) symmetry of AEAs what has attracted more attention to SU($N$)-symmetric Hamiltonians, both theoretically and experimentally~\cite{Gorshkov2010,Cazalilla2014,Takahashi_review,Wu2024review}. In the last decade, a series of theoretical predictions and state-of-the art numerical calculations have been performed on the SU($N$) FHM for different values of $N$ at $T=0$, finite temperature, filling fractions, and for various geometries and limits.  In all of these cases, the SU($N$) lattice models are predicted to display a variety of interesting phases with novel and rich properties depending on the value of $N$. For instance even at weak-to-intermediate coupling, the SU($N$) FHM can host phases other than the SU($N$)-Fermi liquid that we have discussed in this paper. An interesting example of this is the staggered flux phase, where the lattice translation symmetry is spontaneously broken; this phase naturally arises in the large $N$ limit. \\

Furthermore, the strong-coupling limit of the FHM is also of great interest. In particular, at $1/N$-filling (where $U\gg t$ and there is one particle per site on average). In this case, the system is described by the SU($N$) Heisenberg model:
\begin{equation}
    H =\frac{2 t^2}{U} \sum_{\langle i, j \rangle} \sum_{\alpha,\beta} S_{\alpha}^{\beta} (i)  S_{\beta}^{\alpha} (j),
\end{equation}
where the spin operators, $S_{\alpha}^{\beta} = c_{\alpha}^{\dagger} c^{\phantom{\dagger}}_{\beta}$ now obey the SU($N$) algebra. This model can host a zoo of exotic phases in two dimensions. For instance, at large N, it can host the non-magnetic lattice-symmetry breaking valence cluster state, and the chiral spin liquid which supports topological fractional excitations~\cite{Hermele2011}. Interestingly, this model can support ordered states at smaller $N$ for other lattice geometries. For instance, the SU(3) Heisenberg model hosts a three-sublattice state in a triangular lattice~\cite{Toth2010,Bauer2012} and a dimerized magnetically ordered state on a honeycomb lattice~\cite{corboz2013competing,zhao2012plaquette,lee2012spontaneous}. Finally, we note that in a recent study, Yamamoto {\it et al.} have demonstrated the existence of nematic phases in the SU(3) Heisenberg model on a triangular lattice in the presence of low and high magnetic fields~\cite{Yamamoto2020}. \\

In this review, we chose to focus on the experimental achievements of quantum simulators with AEAs in optical lattice. However, there is an abundant amount of theoretical work on SU($N$) models and we have highlighted only a few important works in this section. We thus believe that it would be instructive to provide a short compendium of theoretical and computational work that has been done for the SU($N$) lattice models. This is done in~\Tref{numericalwork} where references on computational and theoretical work are presented. These works illustrate the richness of the models and motivate further experimental studies.

\section{Future directions}
\label{sec:Future}

With the development and implementation of quantum gas microscopy for AEAs~\cite{Yamamoto2016,Okuno2020,Buob2024}, number-resolved imaging without parity projection~\cite{Su2024}, and cooling proposals specific to SU($N$) gases~\cite{Merlin2021,Yamamoto2024}, experiments now poise themselves to measure long range correlations and explore the plethora of proposed magnetic states (see Refs. in~\Tref{numericalwork}). Furthermore, a key objective in many-body physics is to understand the effects of doping Mott insulators and magnetically ordered phases. This has been an active area in the past few years for SU(2) ultracold FHMs. One example of progress is the development and testing of geometric string theory~\cite{Chiu2019,Koepsell2020}, which establishes a connection between the strongly correlated quantum states at finite doping and the AFM parent state at half-filling. Even at the current temperatures, experiments are equipped to explore the doped SU($N$) FHM and study the string length and anisotropy across different magnetic crossovers at finite temperature. It is worth mentioning that a first step in this direction for the SU($N$) FHM was taken in Ref.~\cite{Henning2023}. \\

Exploring the spin-imbalanced SU($N$) FHM and the SU($N$) FHM under symmetry-breaking fields is an immediate question of interest to the field. These questions can be addressed in a straightforward manner by creating spin-polarized samples via optical pumping. In this context, we note that He {\it et al.} have recently investigated the effect of spin imbalance on the thermodynamics of SU($N$) Fermi gases~\cite{He_2024}. Other possible directions of study correspond to the experimental exploration of Nagaoka ferromagnetism~\cite{Botzung2024b,Rajiv2022c}, the ionic Hubbard model~\cite{Wang2022}, and the SU($N$) FHM with Raman couplings, which is predicted to support chiral currents~\cite{Ferraretto2023}. In addition to SU($N$) quantum magnetism, other interesting directions which could be readily implemented with AEAs in OLs, correspond to the study of transport properties, (such as was done for SU(2) in Refs.~\cite{Brown2018,Nichols2018}), engineering hadronic matter~\cite{Werner2023}, realizing collective spin models~\cite{jakab2021quantum}, and the quantum simulation of lattice gauge theories~\cite{surace2024scalable}. As previously mentioned, the interest in using AEAs in OLs  is not limited to studying the SU($N$) FHM model, but also two-band models~\cite{Sotnikov2020,Gorshkov2010,Takahashi_review} such as the SU($N$) Kondo Lattice Model (KLM)~\cite{Coqblin1969,Doniach1977,Tamura2023,Totsuka2023}. These two-band models can be implemented by using the $^1S_0$ ground state and the $^3P_0$ metastable state of AEAs, which exhibit SU($N$) symmetric interactions.\\

Finally, these systems can be a versatile arena to study far-from-equilibrium physics. There have already been a few theoretical efforts to study the effects of quantum quenches in these systems~\cite{werner2020quantum,zhang2019quantum,feher2020generalized,mamaev2022resonant,Huang2020}. These systems can be potentially employed to study generalized thermalization~\cite{ilievski2015complete}, and hydrodynamics~\cite{ruggiero2020quantum} dynamical quantum phase transitions~\cite{heyl2018dynamical}, and the physics of non-thermal fixed points~\cite{mikheev2023universal}. In this context, it is worth noting that although interactions between fermionic AEAs are inherently repulsive ($U>0$), and are not tunable via magnetic Feshbach resonances as in the case of alkali atoms, the possibility to study attractive ($U<0$) SU($N$) FHMs with $N$ as large as $36$ using ultracold molecules has been recently proposed~\cite{Mukherjee2024}. This opens up the possibility of performing interaction quenches in these systems. It would also be interesting to implement periodic driving in these systems~\cite{chinzei2020time}, thereby paving the path towards realizing Floquet phases of matter with SU($N$)-symmetric interactions.\\
 
\section*{Acknowledgements}

EIGP is supported by the grant DE-SC-0022311, funded by the U.S. Department of Energy, Office of Science. SC thanks DST, India for support through SERB project SRG/2023/002730. We thank Kaden Hazzard for several discussions.

\section*{References}
\bibliographystyle{unsrt}
\bibliography{SUN_lattice.bib}

\end{document}